\pdfoutput=1
\documentclass[manuscript]{acmart}
\usepackage{graphicx}
\usepackage{booktabs}
\usepackage{caption}
\usepackage{subcaption}
\usepackage{subfig}
\usepackage{listings}
\usepackage{etoolbox}
\AtBeginDocument{%
  }

\makeatletter
\@ifundefined{sf@counterlist}{\let\sf@counterlist\@empty}{}
\makeatother
\begin{document}
\hypersetup{pdfauthor={}, pdftitle={SherpaAI: A Multi-modal Solution for Delivering Personalized and Adaptive Fitness Interventions}}

\title{SherpaAI: A Multi-modal Solution for Delivering Personalized and Adaptive Fitness Interventions}

\author{Shivangi Agarwal}
\authornote{These authors contributed equally to this research.}
\email{shivangi.agarwal@plaksha.edu.in}
\affiliation{%
  \institution{HTI Lab, Plaksha University}
  \city{Mohali}
  \country{India}
}

\author{Zoya Ghoshal}
\authornotemark[1] 
\email{zoya.ghoshal@plaksha.edu.in}
\affiliation{%
  \institution{HTI Lab, Plaksha University}
  \city{Mohali}
  \country{India}
}

\author{Bharat Jain}
\authornotemark[1] 
\email{bharat.jain@plaksha.edu.in}
\affiliation{%
  \institution{HTI Lab, Plaksha University}
  \city{Mohali}
  \country{India}
}

\author{Siddharth}
\email{siddharth.s@plaksha.edu.in}
\affiliation{%
  \institution{HTI Lab, Plaksha University}
  \city{Mohali}
  \country{India}
}

\renewcommand{\shortauthors}{Anon.}

\begin{abstract}
Personalization of exercise routines is a crucial factor in helping people achieve their fitness goals. Despite this, many contemporary solutions fail to offer real-time, adaptive feedback tailored to an individual's physiological states. Contemporary solutions often rely only on static, pre-set plans and rarely adjust in real time to factors such as a user's pain thresholds, fatigue levels, or form during a workout. This work introduces SherpaAI, a multi-modal system that unifies computer vision, physiological sensing (heart rate and voice), and the reasoning capabilities of Large Language Models (LLMs) — modalities that prior systems have largely explored in isolation — to deliver real-time and individually-adaptive guidance across a set of strength, balance, and flexibility exercises. SherpaAI continuously monitors a user's physical form and level of exertion, among other parameters, to provide dynamic interventions focused on exercise intensity, rest periods, and motivation. To validate our system, we performed a technical evaluation confirming our models' accuracy and quantifying pipeline latency, alongside an expert review where certified trainers validated the correctness of the LLM's interventions. Furthermore, in a controlled within-subject study with 25 participants, SherpaAI demonstrated significant improvements over a non-adaptive baseline modeled on typical self-guided workouts (e.g., following online videos or general-purpose AI chat tools for guidance). With SherpaAI, users reported significantly greater enjoyment, a stronger sense of achievement, and significantly lower levels of boredom and frustration. These results indicate that by integrating multi-modal sensing with LLM-driven reasoning, adaptive systems like SherpaAI can create a more engaging and emotionally satisfying workout experience. Our work provides a blueprint for integrating multi-modal sensing with LLM-driven reasoning, demonstrating that it is possible to create adaptive coaching systems that are not only more engaging but also safe and reliable within a controlled setting.

\end{abstract}


\begin{teaserfigure}
  \includegraphics[width=\linewidth]{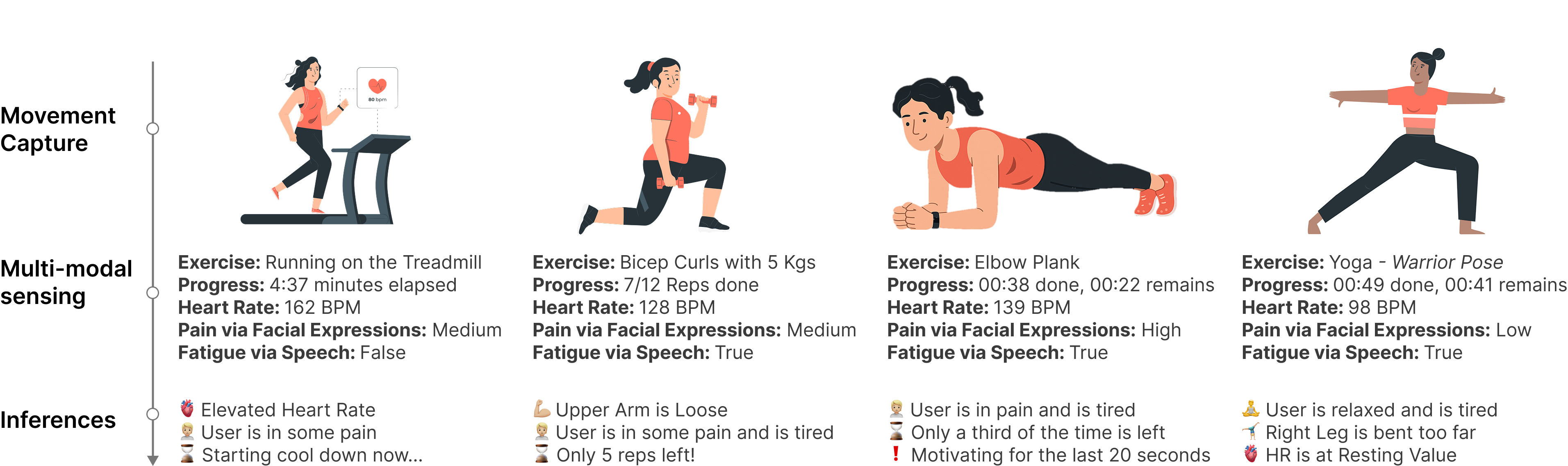}
    \caption{\textbf{Introducing SherpaAI, a personalized, multi-modal AI fitness assistant designed to enhance workout efficiency by providing real-time adjustments to users. It (A) captures physiological data through sensors, (B) interprets physical and emotional states during workouts, and (C) delivers tailored guidance to optimize exercise form, intensity, and safety. SherpaAI ensures workouts remain effective while respecting individual limitations and promoting proper technique.}}
  \label{fig:teaser}
\end{teaserfigure}



\maketitle

\section{Introduction}
Working out is an essential part of our daily lives, yet finding guidance that truly understands our body and its specific limitations can be challenging. Personalized attention from physical trainers often leads to safer, more effective workouts that show consistent progress. However, employing personal trainers, while ideal, can be prohibitively expensive or inaccessible for many individuals.


The fitness technology landscape has shifted significantly toward personalization and adaptiveness taking the center stage, further propelled by growing consumer demand \cite{huang2024exploring, info:doi/10.2196/55964}. In response, research has emphasized developing adaptive systems that generate workout plans using user-specific health metrics \cite{bays2022brief, novatchkov2013artificial, shin2023planfitting} and recommend daily routines based on physiological indicators \cite{barber2017feasibility, lee2024multimodal}. This brings out the need for a continuously evolving approach to personalize fitness routines, recognizing unique patterns and refining recommendations to maximize impact.


Building on these emerging capabilities, earlier efforts to replicate the adaptability of human trainers primarily focused on fundamental exercise elements, such as pose estimation for form correction and basic technique feedback \cite{kanase2021pose, kwon2022real,6199869, 10.1145/2851581.2892519}. Previous works have employed bio-sensing technologies to capture indicators like heart rate variability, breathing patterns, and muscle fatigue, thereby enhancing the potential for individualized feedback \cite{henriksen2018using, qiu2017survey, passos2021wearables}. These advances, however, remain as separate modules rather than highly cohesive systems. As a result, earlier technologies often rely on generic recommendations that still lean heavily on basic metrics such as height, weight, or Body Mass Index (BMI) \cite{4575013, 10.1145/2769493.2769507, mekruksavanich2022multimodal}, which fail to capture the nuanced demands of individual users. This shortfall highlights a pressing research gap: the need for adaptive fitness solutions that go beyond just addressing elementary demographic variables. Instead, these systems must dynamically learn and adjust in the moment, delivering context-aware guidance that mirrors the responsiveness of a personal trainer and fostering long-term engagement in safe, effective workouts.

With the advent of Large Language Models (LLMs), the fitness domain is poised to evolve beyond simple customizations \cite{li2022artificial, hassoon2021randomized, weemaes2024effects, info:doi/10.2196/22845}. While LLMs are currently underutilized in fitness technology---primarily limited to providing post-workout advice \cite{kim2024health, 10.1145/3613904.3642032}---they present an unprecedented opportunity to create enjoyable and adaptive fitness experiences. By integrating LLMs with computer vision and bio-sensing, future fitness technology can interpret complex physiological and emotional data to create truly personalized, adaptive interventions. This AI-powered ecosystem can evolve with an individual's progress and goals, that transforms exercise experiences and helps people reach their fitness potential.


In this paper, we present SherpaAI, a system that demonstrates a novel approach to adaptive personalization in fitness. SherpaAI integrates multi-modal sensing---including computer vision for movement analysis, facial expression recognition for pain, microphones for vocal fatigue, and heart rate monitoring---with a hierarchical LLM-based reasoning module. The system is designed to interpret these real-time physiological and biomechanical inputs to provide tailored interventions on form, intensity, and motivation. We leverage LLMs not as passive, static advisors but as adaptive collaborators that interpret physical exertion indicators and generate metrics uniquely tuned to each user's emotional state, motivational preferences, and fitness goals.

We started with conducting a formative study with 90 participants to understand the scope of creating a multifarious and personalized AI fitness coach. From our findings, we concluded that personal trainers were certainly not a norm, and most would not even consider hiring them. Most participants were very open to the idea of having an AI assistant to guide them through a fitness routine, and a majority of them wanted features like personalized workout plans, progress tracking, performance analytics, and health data integration (incorporating factors like sleep and nutrition). This  process allowed us to identify key system problems and clarify customer expectations.


These insights thus informed the development of SherpaAI, an adaptive AI system that can provide feedback on how to rectify form in real-time, modify workout intensity, and push you safely out of your comfort zone. SherpaAI leverages (1) a multi-modal sensory input system combining cameras for capturing movement and facial expressions, smartwatches to capture heart activity, and microphones for user speech (2) a processing module to get insights on various physiological indicators---like physical exertion, pain, and fatigue---to understand a user's state during a fitness routine in real-time (3) the reasoning capabilities of LLMs to utilize these inferences and provide interventions---such as rest period modifications, encouraging messages, intensity adjustments---delivered based on need (4) text-to-speech models to deliver verbal, tone-adaptive feedback to the user through an in-ear assistant. 

In summary, we contribute:
\begin{itemize}
    \item \textbf{Personalized Feedback and Modifications} A system to identify subtle physiological cues to adjust workout intensities, building comprehensive understandings of individual baseline patterns and exertion thresholds through continuous multi-modal analysis.
    \item \textbf{Real-time and Contextually Appropriate Interventions} A seamless guided experience that examines changes in user state as they occur, providing relevant and accurate interventions in real-time.
    \item \textbf{Adaptive Multi-modal Integration} The implementation of SherpaAI, which leverages pose correction, facial expression recognition, heart rate monitoring, audio data analysis, and LLMs to create a system which constantly reconfigures itself to its user's needs. 
    \item \textbf{Experimental Study-based Validation:} A comprehensive and technical evaluation of SherpaAI's performance based on physiological, visual and audio-based data collected from 25 participants as they worked out in real-time, demonstrating its potential to improve users' fitness experiences.

\end{itemize}

\section{Related Works}
Our work builds upon previous research works conducted in the fields of adaptive and personalized fitness coaching, multi-modal sensing in fitness applications, affective computing, and real-time form analysis with posture correction.

\subsection{Adaptive and Personalized Fitness Coaching}


Recent research in adaptive fitness coaching focuses on tailoring experiences to individual needs. For example, Ilukpitiya et al. \cite{ilukpitiya2024ai} mobile app using CNNs for body-type classification and real-time feedback, while Mohan et al. \cite{mohan2020designing} focused on sedentary individuals with an app that uses adaptive goal-setting algorithms to dynamically adjust weekly goals based on user performance. Systematic analyses have confirmed the effectiveness of such AI-driven approaches for physical activity \cite{oh2021systematic}, with applications extending into educational settings like school PE classes through personalized virtual trainers that use Case-based Reasoning (CBR) to match users based on BMI and personal preferences \cite{mokmin2020effectiveness}.



To advance beyond static personalization, systems like FitRec use LSTMs to analyze dynamic fitness data from wearable devices---including heart rate, GPS, and altitude---to provide real-time feedback based on physiological thresholds. Despite these advances, many studies exhibit a self-report bias, heavily relying on user-reported metrics (diet logs) to generate feedback. Additionally, most studies did not consider gender as a parameter in their analyses. These limitations highlight a critical gap: the overdependence on potentially inaccurate self-reporting rather than objective measurements. This points to the need for the multi-modal sensing approaches we explore.


\subsection{Multi-modal Sensing in Fitness Applications}


The evolution of multi-modal sensing in fitness applications has combined data from various sources to monitor performance. Wearables have emerged as foundational tools for gathering physiological data. For instance, FitCoach combined wrist-worn wearables with smartphone sensors to track exercises and interpret motion strength and speed \cite{guo2017fitcoach}. Other work has fused inertial sensors with camera systems for 3D pose detection during specific exercises like barbell squats, demonstrating the value of multi-sensor fusion \cite{wilk2020multimodal}. Directly relevant to our work, Chowdhury et al. \cite{chowdhury2019prediction} created a system utilizing heart rate, electrodermal activity (EDA), and skin temperature with machine learning models to classify exercise intensity.


Despite these advances, personalization still remains an underexplored frontier. While recent systems like UbiPhysio \cite{wang2024ubiphysio} heavily focus on biomechanical form evaluation for rehabilitation, and GPTCoach \cite{jorke2025gptcoach} targets physiological metrics and motivation, most foundational systems still apply standardized metrics across users without adapting to individual biomechanics. Unlike these systems, SherpaAI deeply integrates both real-time pose estimation and continuous physiological and affective LLM reasoning to dynamically adjust form and intensity. Analyzing such works have led us to explore a new dimension in the field of fitness: affective computing as a component in exercise personalization.


\subsection{Real-Time Form Analysis and Correction Systems}


Previous research in real-time correction and feedback has been established as pivotal components of effective solutions, aiming to prevent injuries through timely interventions using AI and computer vision technologies. For example, Kotte et al. \cite{kotte2023real} explored a real-time feedback system for conventional gym exercises using YOLOv7-pose to detect key points and calculate joint angles. This approach has been extended to other domains, such as workplace safety, where AI-driven posture monitoring systems combine MediaPipe landmarks with LSTMs to analyze manual lifting tasks and prevent musculoskeletal disorders (MSDs) \cite{bagga2024real}. Research has also focused on specific, form-focused exercises, with systems providing detailed feedback on joint misalignments in yoga poses \cite{anand2022yoga} or using IoT sensors and KNN classifiers to guide users in exercises like bicep curls \cite{hannan2021portable}.



While these prior works have significantly advanced real-time posture correction and feedback systems, they share common limitations: inadequate generalization across diverse body types, environmental conditions, and exercise variations. This generalization gap is, at its core, a personalization gap — these systems apply the same fixed correction thresholds to every user rather than adapting to an individual's body mechanics, fitness level, or exercise history. Additionally, computational overhead often results in feedback delays that reduce intervention effectiveness.


The research surveyed across adaptive coaching, multi-modal sensing, affective computing, and real-time form analysis demonstrates significant advances in fitness technology, yet reveals a persistent gap in personalization and real-time adaptability. While existing systems excel in isolated domains—whether tracking physiological markers, analyzing emotional states, or correcting posture—they often fail to integrate these elements into a cohesive, responsive system. Our work addresses this gap by combining multi-modal sensing \cite{bays2022brief, novatchkov2013artificial, shin2023planfitting, stromback2020mm, zou2020low}---including microphones for breathing analysis and optical heart rate sensors---with affective computing and real-time form correction through an LLM-based reasoning module. This integrated approach enables our system to simultaneously monitor physiological indicators, detect emotional responses, and provide personalized feedback. By quantifying fatigue through these comprehensive markers and dynamically adjusting workout intensity based on individual thresholds, we create a truly adaptive fitness experience that evolves with the user's changing physical state.

\subsection{Affective Computing in Fitness Contexts}


The integration of affective computing in fitness applications represents an emerging area to enhance personalization by incorporating emotional states and pain detection. For instance, researchers have used CNNs to classify exercise intensity from facial expression analysis during stationary cycling \cite{khanal2019classification}. Others have created facial expression-based perceived exertion (FRPE) scales by correlating visual markers with heart rate data  \cite{chen2017rating, cascella2024employing}. This line of work has identified specific visual biomarkers---such as open mouths, jaw drops, and nose wrinkles---that consistently correlate with high physical exertion, providing valuable insights into users' subjective experiences \cite{bartlett2005recognizing}.

The application of affective computing also extends to pain assessment. Studies have evaluated AI/ML methods for pain detection using both facial analysis \cite{nagireddi2022analysis} and vocal biomarkers like pitch and intensity \cite{borna2023review, nagireddi2022analysis}. A significant limitation across these studies is their reliance on controlled laboratory environments and their focus on single modalities (visual or audio) without integrating comprehensive physiological markers. This creates a critical research gap: the absence of systems that can monitor and respond to the complete physiological state of users in natural exercise conditions. Real-time posture correction addresses this gap by implementing adaptive and visual feedback mechanisms that function in varying environments to create a comprehensive understanding of the user's physical state and movement patterns.


\section{Formative Study}
\begin{figure*}[htbp]  
  \centering
  \captionsetup[subfigure]{font=scriptsize, labelfont=bf, justification=centering, singlelinecheck=false}
  
  \begin{subfigure}[b]{0.33\textwidth}  
    \centering
    \includegraphics[width=\linewidth]{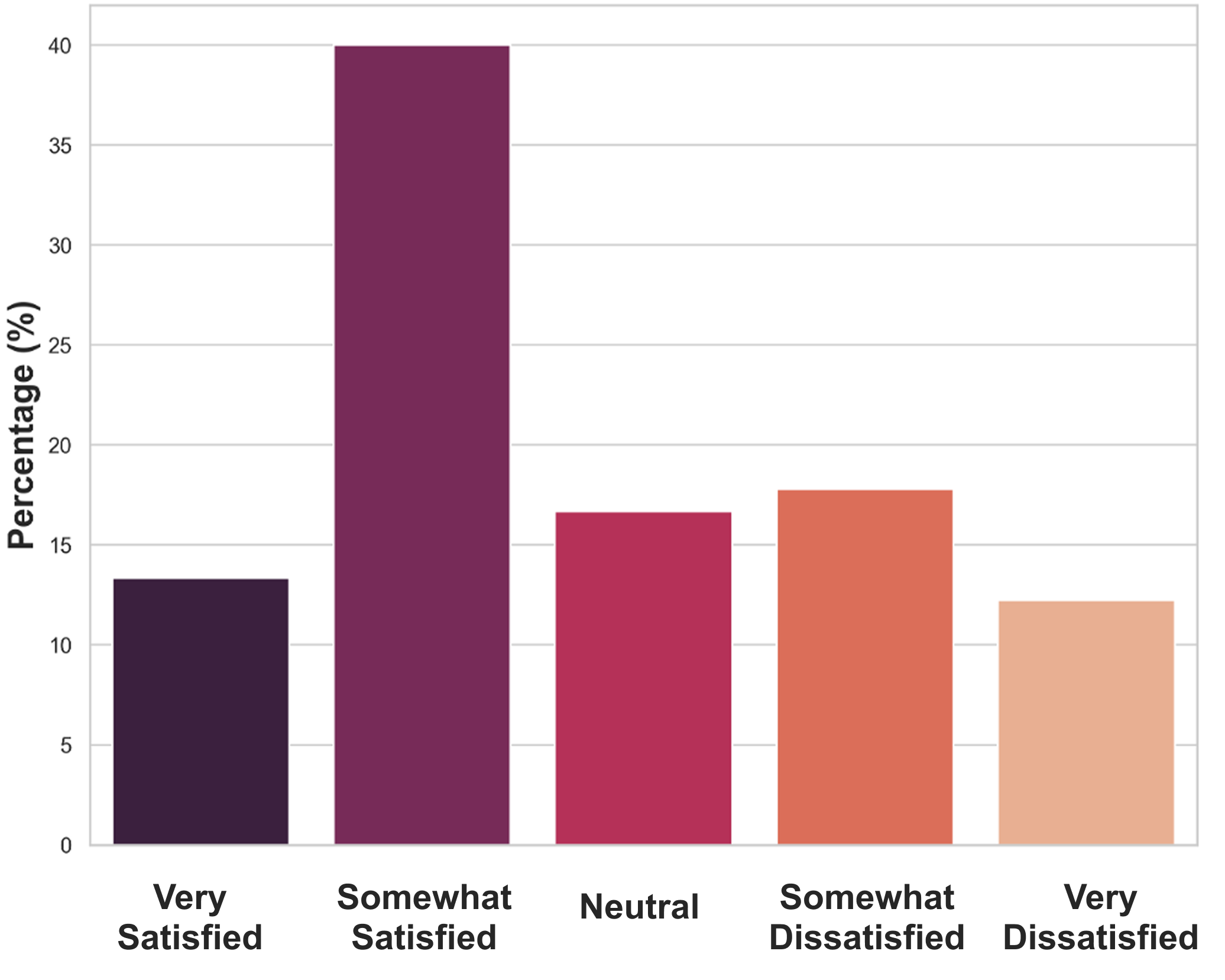}  
    \caption{\textbf{Satisfaction with Current Fitness Routines}}
    \label{fig:satisfaction}
  \end{subfigure}\hfill%
  \begin{subfigure}[b]{0.33\textwidth}
    \centering
    \includegraphics[width=\linewidth]{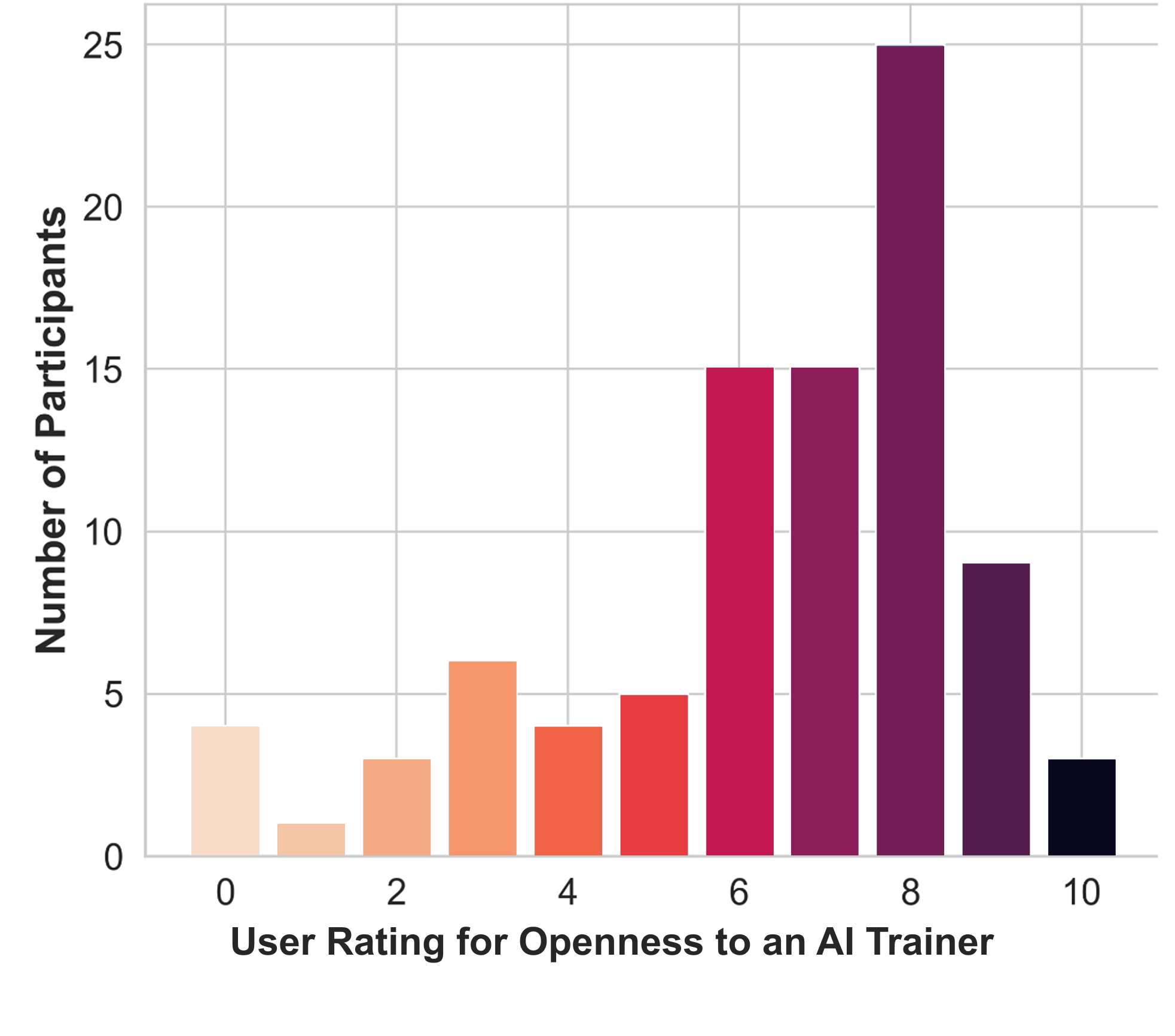}  
    \caption{\textbf{User Receptiveness to an AI Health Coach}}
    \label{fig:likeness}
  \end{subfigure}\hfill%
  \begin{subfigure}[b]{0.33\textwidth}
    \centering
    \includegraphics[width=\linewidth]{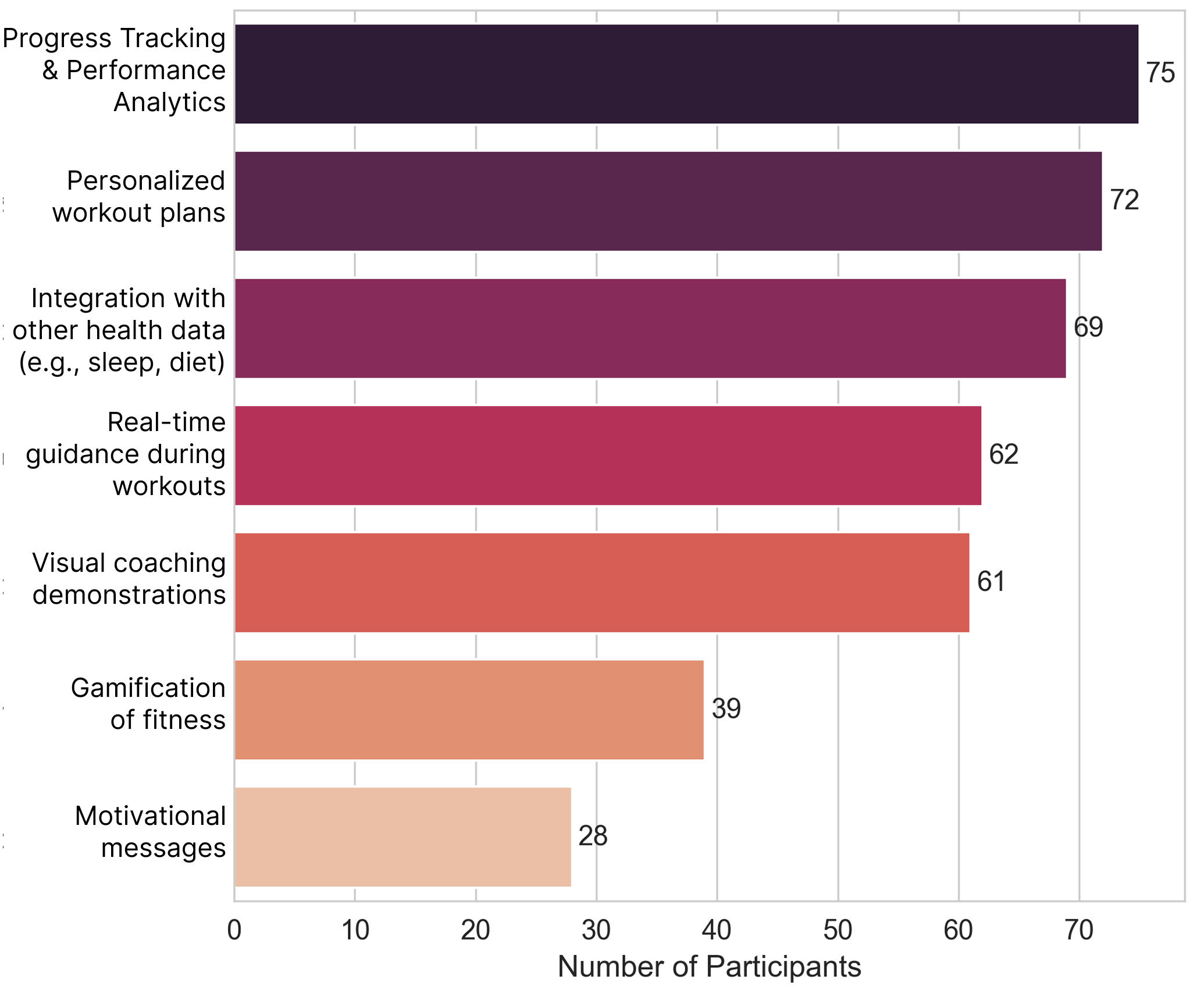}  
    \caption{\textbf{Features Expected from an AI Health Coach}}
    \label{fig:features}
  \end{subfigure}
  
  \caption{\textbf{The distributions illustrate how satisfied users are with current routines, how receptive they would be to an AI health coach, and the kind of features they would expect from a comprehensive AI health coach}}
  \label{fig:3-column-fig}
\end{figure*}

To understand user needs and preferences for an AI-powered fitness coaching system, we conducted a formative study using an online survey. We recruited 90 young adults (age range: 17 to 27) through mailing lists and social media channels. Participants engaged in fitness routines at varying frequencies: 46\% exercised 3-5 times a week, 18\% daily, 17\% once or twice a week, 12\% occasionally, and 8\% never. The study helped us formalize current fitness behaviors, challenges, and expectations, which directly informed the user design for SherpaAI. In the context of this study, we defined an 'AI Health Coach' to participants as an automated, interactive system capable of providing real-time exercise tracking, personalized feedback, and physiological monitoring without human intervention.


\subsection{Findings}
Our respondent pool consisted of 54 males (60\%) and 36 females (40\%), who displayed a wide range of fitness habits. While over half of the participants (53.33\%) reported satisfaction with their current routines (Figure ~\ref{fig:satisfaction}), we identified significant barriers to effective fitness. 

The most frequently cited challenges were time management (24.44\%), a lack of motivation \cite{louw2012exercise, 10.3389/fpsyg.2015.00835}, and inadequate knowledge of proper technique. We also found that the adoption of professional guidance was low; most participants did not use fitness tracking technology, and only 22.22\% consulted with trainers, citing high cost as the primary barrier \cite{koh2022cross, ferreira2022barriers, nikolajsen2021barriers}. In addition to this, some people displayed reluctance to adopting new-age tools in this field because they disliked the feeling of being controlled by a system which displays incompetence. While this is an obstacle to consider, most people in the fitness domain tend to appreciate the change that AI-powered assistants bring into their lives, often reporting that workout experiences were made enjoyably exhilarating \cite{vietzke2023middle, suo2022influence, james2021mediating}. This divergence indicated a potential opportunity for technology adoption among the interested segment of people.

Participant receptiveness to an AI coach was mixed (Figure ~\ref{fig:likeness}). While a majority were neutral or open to the idea, a significant portion (42.22\%) were skeptical. This skepticism was largely attributed to a lack of trust in AI's reliability, a perceived loss of the ``human element" in coaching, and concerns about ease of use \cite{chin2022increase, terblanche2022comparing}. These findings were critical, as they highlighted our primary design challenge: to build user trust, our system needed to feel credible, contextually aware, and directly responsive to a user's real-time state. This motivated our focus on a multi-modal sensing approach.

When asked about desired features, participants showed a strong preference for progress tracking, personalized workout plans, and health data integration (Figure 2c). In terms of real-time interventions, the most requested features were form correction (70\%), workout modifications based on fitness levels (67\%), and recovery recommendations (66\%) \cite{hassoon2021randomized, li2022artificial, dergaa2024using}.

\subsection{Design Implications}
Based on our findings, we were able to identify several key design implications for developing an effective AI-powered coaching system:

\subsubsection{\textbf{A Balance Between Guidance and Autonomy}}
The varied responses to AI coaching receptiveness indicate that users desire guidance without feeling a complete loss of control. An overly prescriptive system can thus feel restrictive, while a hands-off approach fails to provide value. Therefore, SherpaAI should operate on a principle of sporadic, but targeted intervention. It should allow users to autonomously manage their pace, form, and effort within safe and effective parameters. Guidance should only be triggered when necessary --- such as detecting poor form (which could lead to injury), an unsafe heart rate, counting of repetitions, or the completion of a set. This approach respects user autonomy by not being overbearing, while also building trust by delivering valuable, data-driven assistance precisely when it is needed. 

\subsubsection{\textbf{Real-time Adaptations of Task Difficulty Based on Individual Performance}}
Low satisfaction rates with current methods used in fitness routines points to a need for systems that adapt during a session. SherpaAI should thus go beyond pre-set plans and incorporate dynamic workout adjustments based on live physiological indicators. By continuously monitoring metrics like heart rate, facial expressions of pain, and vocal cues of fatigue, the system should make immediate, data-driven decisions to modify exercise intensity, suggest rest, or alter repetitions to ensure the workout is both challenging and safe. 

\subsubsection{\textbf{Integration of Real-time Physiological Health Metrics for Timely Interventions}}
The high interest in integrating physical well-being indicators (76\%) suggests that users desire an all-inclusive approach, with a system which understands their complete state. SherpaAI should consequently implement a multi-modal sensory system \cite{smuck2021emerging, dunn2018wearables, kaewkannate2016comparison, gay2015bringing} that combines data from cameras (for movement and facial expressions), smartwatches (for heart rate), and microphones (for vocal fatigue).  By fusing these contrasting data streams, the system can build a helpful, real-time profile of a user's pain, fatigue, and physical load, enabling interventions that are more contextually aware than those based on a single data source. 
 
\subsubsection{\textbf{Actionable Analytics and Progress Tracking}}
The strong preference for progress tracking (83\% of participants) suggests users want tangible evidence of improvement. Our audio-first approach requires this to be delivered moment-to-moment rather than through post-workout analytics. SherpaAI should provide real-time progress updates, such as announcing repetition counts and time elapsed or remaining \cite{lynch2020changing, bhargava2020opportunities}. This serves to keep the user informed and engaged throughout the routine. Furthermore, to address the stated need for motivation, the system should deliver timely, encouraging phrases, especially during challenging parts in the set, to boost user perseverance. 

\subsubsection{\textbf{Emphasis on Form Correction and Injury Prevention}}
The strong preference for form correction (70\% of participants) and concerns about proper technique highlighted the importance of this feature. We determined that SherpaAI should implement robust pose estimation for accurate form feedback, provide actionable and clear guidance for rectifying form issues, detect potential injury risks and finally, modify the exercise routine accordingly \cite{jones1999physical, jones2017impact, lisman2017systematic, farley2020relationship}. 

\begin{figure*}[!htp]
    \centering
    \includegraphics[width=0.85\linewidth]{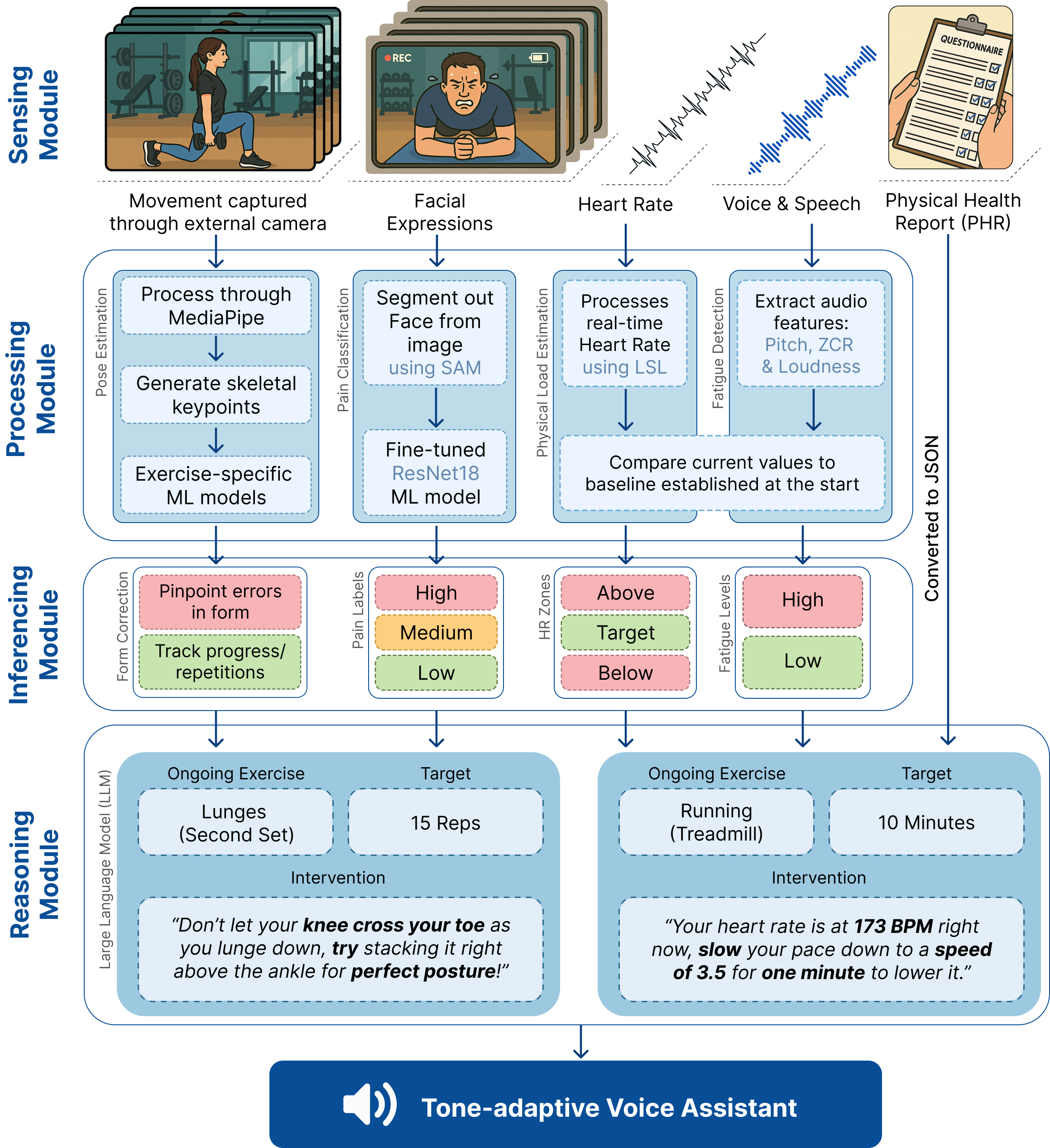}
    \caption{\textbf{SherpaAI's architecture leverages: (1) a sensing module which comprises of cameras, smartwatches, and microphones (2) a processing module which processes sensor data to assess form, pain, physical load, and fatigue (3) an inferencing module which obtains pain labels, and HR and fatigue levels (4) a reasoning module which then leverages LLMs to provide real-time exercise corrections and intensity adjustments (5) a tone-adaptive voice assistant which can deliver in-ear feedback.}}
    \label{architecture}
\end{figure*}

\begin{figure*}[!htp]
    \centering
    \includegraphics[width=0.85\linewidth]{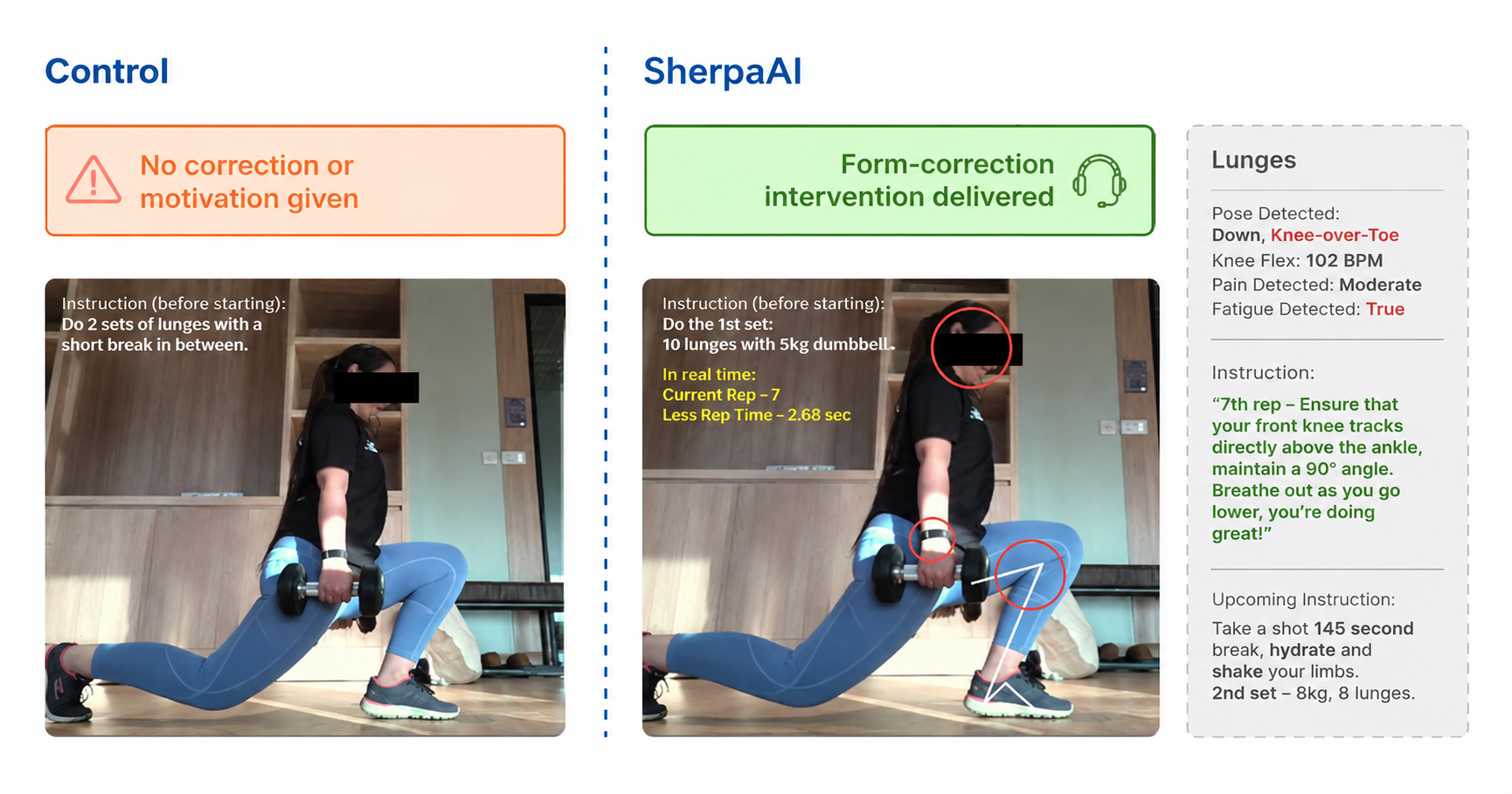}
    \caption{\textbf{Demonstration of how the Control and SherpaAI systems differ from each other, in terms of form correction, repetition counting, and motivational phrases.}}
    \label{control_vs_SherpaAI1}
\end{figure*}

\section{Our Solution: SherpaAI}

SherpaAI has been developed as a personalized AI health coach that leverages multi-modal bio-sensing to build a detailed picture of the user's physical state to provide interventions during the workout in real-time for exercise scheduling, along with form correction, motivation, and adjusting the intensity. SherpaAI allows people to make their workout regimen more effective and push beyond their perceived boundaries. SherpaAI's architecture is shown in Figure ~\ref{architecture}. 
\subsection{Data Capture}

\subsubsection{\textbf{Sensory Input}}

We employ a comprehensive array of sensors to capture critical information about the user's physiological and biomechanical state during workout sessions. The system utilizes two high-resolution external cameras, a smartwatch, and a wireless headset. The specifications and functions of these sensors were:
\begin{itemize}
    \item The cameras record in full HD 1080p resolution with 60 frames per second. One is dedicated to full-body motion tracking and posture analysis, and another is specialized for facial expression recognition and monitoring signs of exertion.
    \item A smartwatch (MAX-HEALTH-BAND by Analog Devices) continuously monitors cardiovascular metrics including heart rate, heart rate variability, and step count.
    \item Additionally, a wireless low-latency headset facilitates both real-time audio communication and binaural feedback while capturing verbal cues and breathing patterns for analysis. This multi-modal sensing approach enables our platform to construct a holistic profile of the user's performance and physiological response to exercise stimuli.
\end{itemize}

\subsubsection{\textbf{Physical Health Report}}
The Physical Health Report (PHR) is created for each user at program initiation to establish their health profile. Users complete the WHO's Global Physical Activity Questionnaire (GPAQ) \cite{gpaq}, which measures time spent on various activities. This is converted to Metabolic Equivalent (MET) \cite{MET} scores, categorizing users as Sedentary, Active, or Very Active.
The PHR also collects metrics like:
\begin{itemize}
    \item Height and weight.
    \item Preferred workout intensity (low, moderate, high).
    \item Previous injuries.
    \item Workout goals (cardiovascular endurance, weight maintenance/loss, muscle gain, flexibility/mobility, or custom).
\end{itemize}
This information then generates a JSON object, with only relevant data passed to each personalized intervention.

\subsection{Processing Module}

We employ a multi-sensory approach along with computer vision techniques to capture movement, pain, fatigue, and heart rate data in real-time. This raw information is then translated into higher-level insights through an inferencing pipeline, which serves as input to an LLM-driven reasoning module. 

\subsubsection{\textbf{Movement-based Pipeline}}
We use computer vision and MediaPipe-based models \cite{lugaresi2019mediapipe} to capture various movement-related details, such as tracking exercise progress and detecting form errors. A continuous video stream is established, and MediaPipe landmarks are generated for each frame in real-time. These landmarks are passed through AI models tailored to the ongoing exercise to monitor proper form. As required, interventions are triggered based on these insights.

\subsubsection{\textbf{Pain}}
Beyond movement data, we also analyze emotional and physiological indicators to gauge pain levels. A second camera records the user’s facial expressions during the workout. This photograph is passed through SAM (Segment Anything Model) \cite{kirillov2023segment} and the face was extracted out. This is then passed to a pain classification model to categorize the user's pain as \textit{High, Medium }or\textit{ Low}.

We developed an ML model, built with pretrained ResNet-18 weights \cite{he2016deep, li2020deep}, and fine-tuned it on the Delaware Pain Dataset \cite{mende2020delaware}. The model classified pain levels into three categories (low, moderate, and high) with 79.3\% accuracy. Thus, our AI framework informed interventions in cardio, strength, and balance modules based on a user's pain levels.

\subsubsection{\textbf{Fatigue}}
Speech and voice features have also been employed to determine user fatigue. During exercise, changes in breathing, pitch, loudness, and Zero-Crossing Rate (ZCR) are good indicators of exertion \cite{rabiner1978digital}. Baseline values were measured at the start of a session; deviations beyond an empirically determined threshold (50–80\%) triggered a ``True'' fatigue state. Fatigue insights helped prompt appropriate suggestions or motivational support.

\subsubsection{\textbf{Heart Activity}}
To track heart rate (HR) in real-time, we integrate a Max-Health-Band device via the Lab Streaming Layer (LSL) \cite{LSL}. The raw HR data is stored in a CSV file alongside timestamps. The baseline HR was the mean of the first 60 readings taken before the session, while the mean of the last 5 readings was used to represent the user’s current HR. These values help us determine safe intensity levels, prompt interventions, and tailor rest periods.

\begin{figure*}[ht!]
  \centering
  
  \begin{subfigure}[b]{0.48\linewidth}
    \centering
    \includegraphics[width=\linewidth]{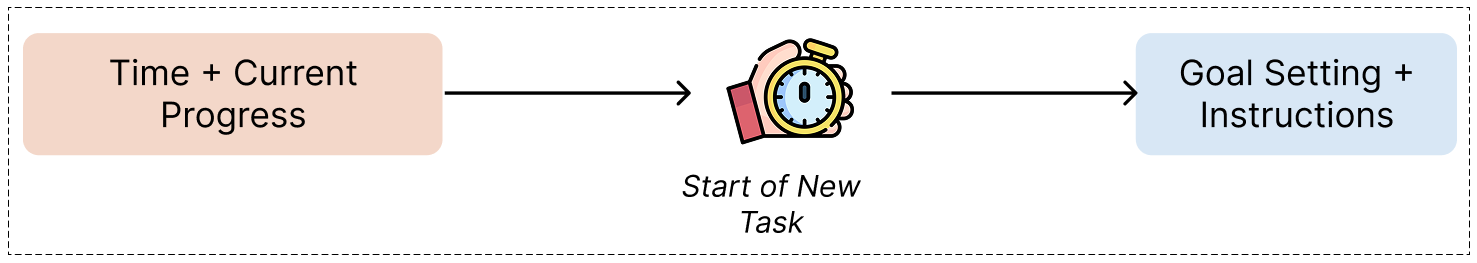}
    \caption{Goal Setting \& Instructions Intervention}
    \label{fig:goal-setting}
  \end{subfigure}
  \begin{subfigure}[b]{0.48\linewidth}
    \centering
    \includegraphics[width=\linewidth]{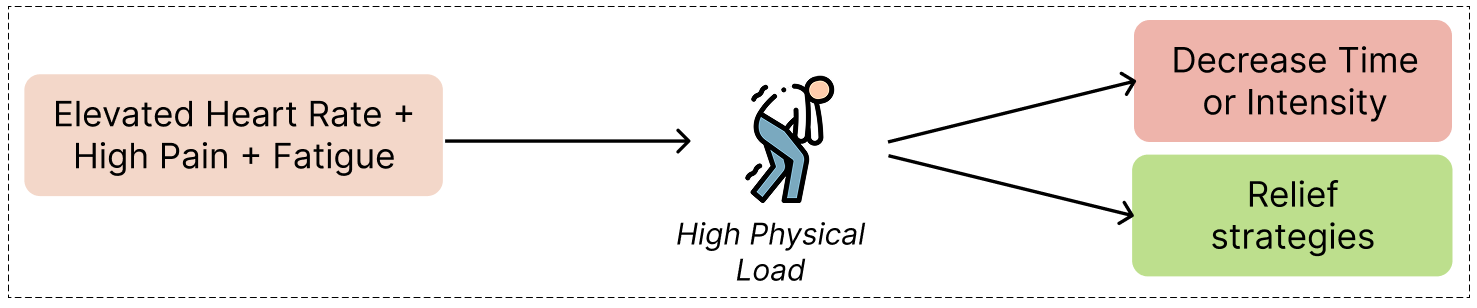}
    \caption{Intensity \& Relief Interventions}
    \label{fig:intensity-relief}
  \end{subfigure}

  \vskip\baselineskip 
  
  \begin{subfigure}[b]{0.48\linewidth}
    \centering
    \includegraphics[width=\linewidth]{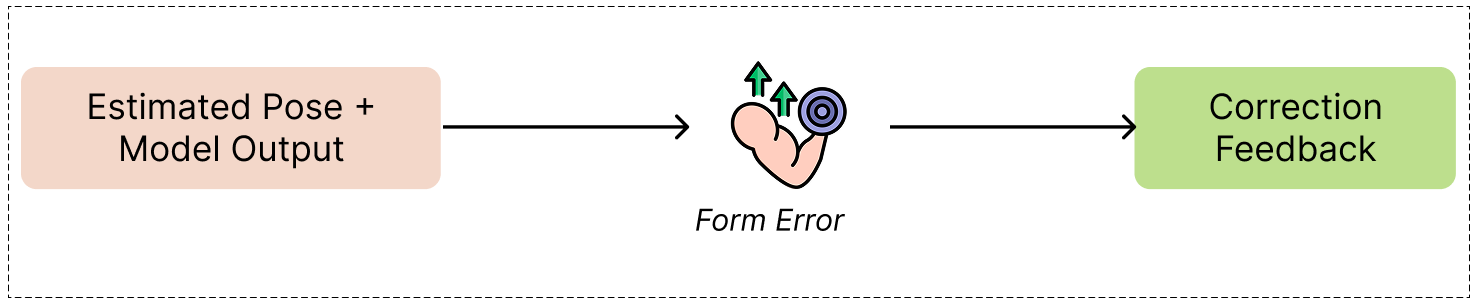}
    \caption{Correctional Feedback Intervention}
    \label{fig:correct-form}
  \end{subfigure}
  \begin{subfigure}[b]{0.48\linewidth}
    \centering
    \includegraphics[width=\linewidth]{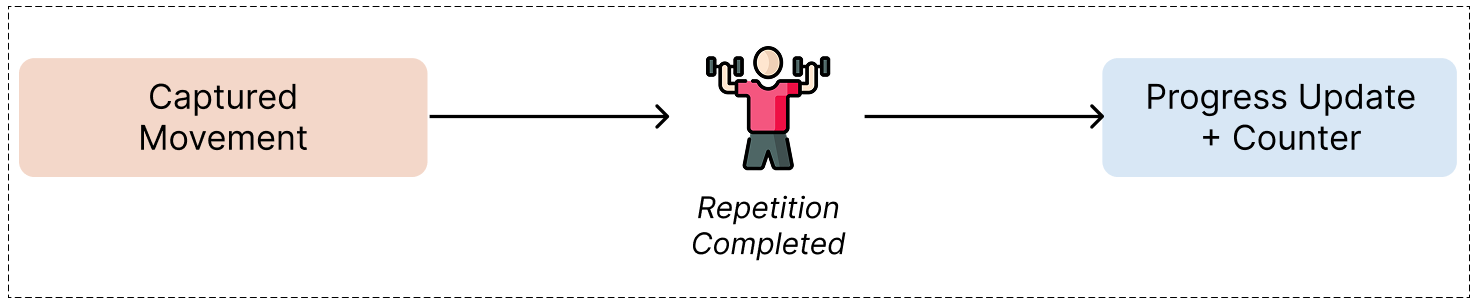}
    \caption{Progress Update \& Counter Intervention}
    \label{fig:progress-counter}
  \end{subfigure}
  
  \vskip\baselineskip 

  \begin{subfigure}[b]{0.48\linewidth}
    \centering
    \includegraphics[width=\linewidth]{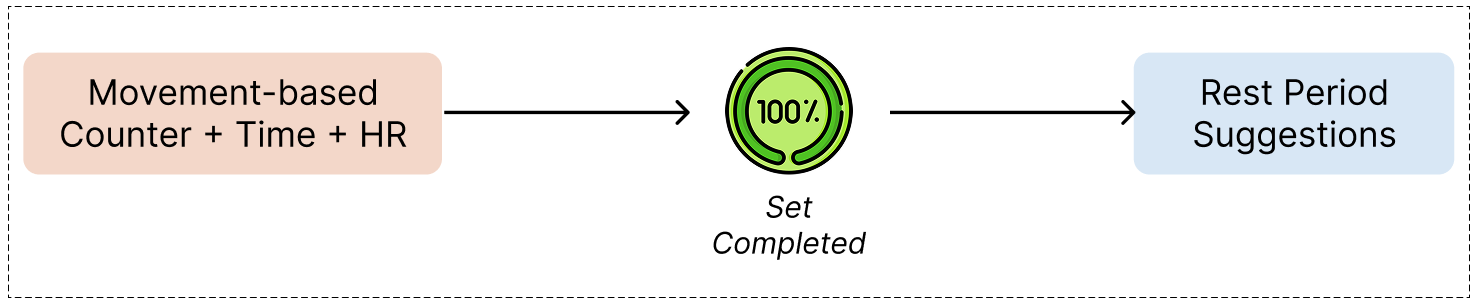}
    \caption{Rest Period Suggestions Intervention}
    \label{fig:rest-period}
  \end{subfigure}
  \begin{subfigure}[b]{0.48\linewidth}
    \centering
    \includegraphics[width=\linewidth]{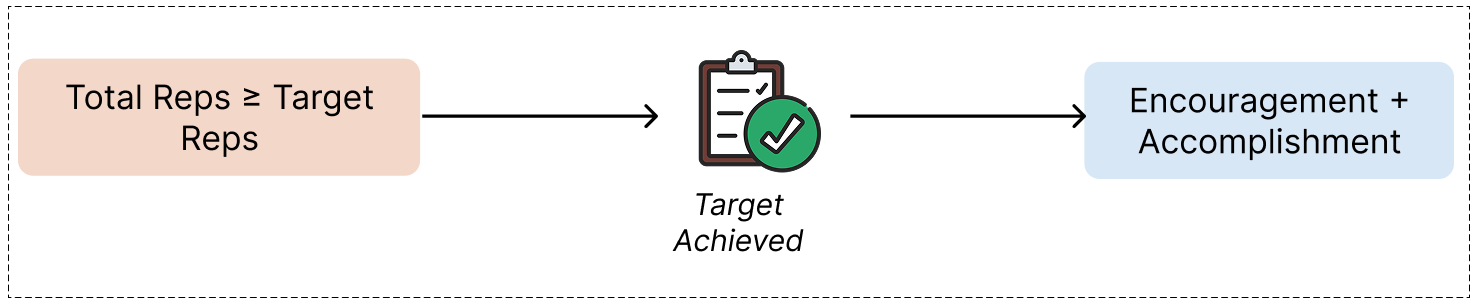}
    \caption{Milestone/Accomplishment Encouragement Intervention}
    \label{fig:encouragement}
  \end{subfigure}

  \vskip\baselineskip 

  \begin{subfigure}[b]{0.96\linewidth} 
    \centering
    \includegraphics[width=\linewidth]{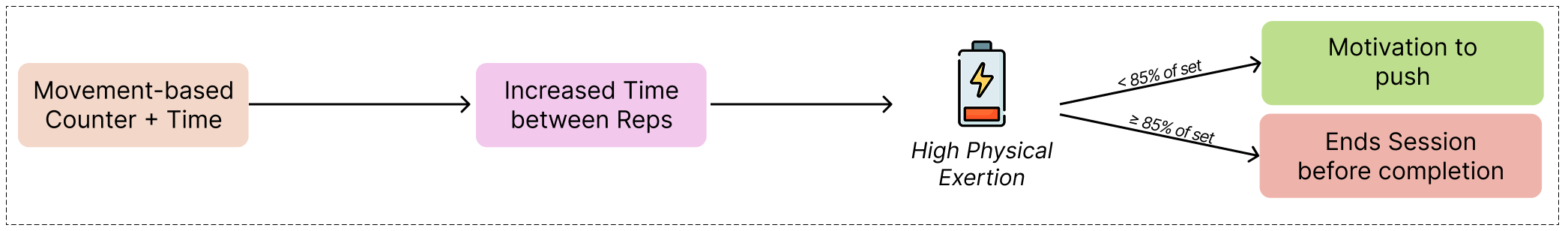}
    \caption{End Motivation \& Session Interruptions Intervention}
    \label{fig:motivation-interruptions}
  \end{subfigure}
  
  \caption{\textbf{Specific triggers lead to their corresponding interventions during an exercise routine including form correction feedback, goal setting, intensity adjustments, rest suggestions, encouragement of accomplishments and milestones, progress updates, and repetition counting announcements.}}
  \label{fig:interventions}
  \Description{}
\end{figure*}



\subsection{Inferencing Module}
The primary function of the Inferencing Module is to convert the clean numerical features from the Processing Module into a set of meaningful, categorical labels. It applies domain-specific logic, biomechanical rules, and thresholds to make judgments about the user's performance and physiological state. These labels are then aggregated into a single structured JSON object that served as the real-time input for the LLM-driven Reasoning Module.

\subsubsection{Motivation for Exercise Selection}
We focus on four fundamental fitness routines namely lunges, bicep curls, elbow planks, and basic yoga poses (\textit{tree, warrior, downward-facing dog}) for several reasons. First, these exercises target major muscle groups commonly emphasized in standard fitness guidelines \cite{riebe2018acsm, bull2020world}. Second, they are relatively straightforward to perform and observe, making them well-suited for computer vision-based analysis \cite{smith2010model}. Finally, each exercise addresses a distinct pillar of fitness---endurance, strength, balance, and flexibility---ensuring that our system remains comprehensive for a wide range of users \cite{johnson2007exercise,lee2008twelve}. Interventions during the exercise routine were provided as per Figure ~\ref{fig:interventions}.

\paragraph{\textbf{Lunges}}
Our system recognizes different stages of a lunge---initial, middle, and down---allowing for a repetition counter to monitor within-set progress. We also track the time between two consecutive lunges as an indicator of user exertion, with interventions for the same being conveyed as per Figure ~\ref{fig:motivation-interruptions}. Additionally, SherpaAI checks for the ``knee-over-toe'' error to ensure that the forward knee stacks directly above the ankle, distributing weight properly and reducing strain on the knee joint.

\paragraph{\textbf{Bicep Curls}}
Similar to lunges, a repetition counter tracked within-set progress. SherpaAI provides two main form corrections: (1) “loose upper arm,” triggered when the user moves the upper arm instead of hinging at the elbow, and (2) “weak peak contraction,” triggered when the user is not fully engaging the biceps at the top of the movement. Interventions are provided as per Figure ~\ref{fig:correct-form}. Correcting these errors promotes optimal bicep engagement and reduces shoulder or forearm strain.

\paragraph{\textbf{Elbow Plank}}
Each video frame is classified as either “high back,” “correct pose,” or “low back.” The timer for the plank began only when the correct form is detected. Users received actionable steps to correct their posture if errors occur, improving exercise efficiency and reducing strain on the shoulders and back.

\paragraph{\textbf{Yoga}}
For yoga poses (\textit{tree, warrior, and downward-facing dog}), we compute joint angles using MediaPipe landmarks to detect correct or incorrect positioning. These angles and flagged joints guide user feedback for alignment and adjustments, ensuring safer and more effective practice.

\paragraph{\textbf{Physiological State Inference}}
In parallel, the module inferred the user's physiological state from other sensor streams.
\begin{itemize}
    \item \textbf{Pain:} The probability distribution from the pain classification model is converted to a discrete label (e.g., `High', `Medium', `Low') by selecting the class with the highest confidence.
    \item \textbf{Fatigue:} The vocal feature deviations (pitch, loudness, ZCR) are compared against a threshold. Based on pilot testing with 5 users, we determined that changes greater than 60\% from baseline were a reliable indicator of self-reported fatigue, triggering a `true' fatigue state.
    \item \textbf{Heart Activity:} The user's current BPM is compared against their target heart rate zone. This zone is calculated for each user via the Karvonen Method:
    $$ {Target\ HR} = (({Max\ HR}\ -\ {Resting\ HR})\ \times \ \% \ {Intensity})\ +\ {Resting\ HR} $$
    Based on this comparison, the module inferred whether the user's heart rate is `Above', `Target', or `Below' the desired zone.
\end{itemize}


\subsection{Reasoning Module}
The Reasoning Module uses a hierarchical, task-dependent prompting strategy to leverage the LLM for different cognitive tasks during a workout. A single, generic prompt is insufficient for both high-level planning and low-level real-time feedback. Instead, we employed two distinct prompt structures:

\begin{enumerate}
    \item \textbf{Inter-Exercise Transition Prompt:} Used for macro-level planning between different workout modalities (e.g., cardio and strength). This prompt asks the LLM to act as a planner, synthesizing the user's PHR and recent physiological data to determine optimal rest periods or adjustments for the next phase of the workout.
    \item \textbf{Intra-Exercise Intervention Prompt:} A real-time prompt used for micro-level feedback during an exercise. It is optimized for low latency and provides the LLM with a snapshot of the user's immediate state to generate concise, actionable feedback on form, intensity, or motivation.
\end{enumerate}

Figure ~\ref{fig:evolved_prompt} shows simplified examples of both prompt types, illustrating how the LLM's task is framed differently based on the context.

\begin{figure}[h!]
\centering
\caption{\textbf{Examples of our hierarchical prompting strategy. (a) An Inter-Exercise prompt for planning rest. (b) An Intra-Exercise prompt for real-time form correction.}}
\label{fig:evolved_prompt}
\begin{subfigure}[t]{\linewidth}
\centering
\begin{lstlisting}[basicstyle=\small\ttfamily, breaklines=true]
----------PROMPT (a): Inter-Exercise Transition----------
You are a personalized AI fitness coach. Determine an appropriate rest period.
Current data:
- Just completed: Cardio
- Next exercise: Lunges
- Baseline heart rate: 65 bpm
- Current heart rate: 145 bpm
- Physical Health Report: 
\end{lstlisting}
\begin{lstlisting}[basicstyle=\small\ttfamily, breaklines=true]
{"fitness_level": "Active", "goal": "endurance"}
\end{lstlisting}
\begin{lstlisting}[basicstyle=\small\ttfamily, breaklines=true]
Constraints:
- Max rest: 60 seconds.
- Adjust based on HR elevation and intensity transition.
- Output JSON with "seconds" and an encouraging "message".
\end{lstlisting}
\caption{Inter-Exercise Transition Prompt}
\label{fig:inter_exercise}
\end{subfigure}
\vspace{1em}
\begin{subfigure}[t]{\linewidth}
\centering
\begin{lstlisting}[basicstyle=\small\ttfamily, breaklines=true]
----------PROMPT (b): Intra-Exercise Intervention----------
You are SherpaAI, a concise fitness coach. Based on the user's real-time state, 
provide a short, direct intervention (max 15 words).
Real-time state (JSON):
\end{lstlisting}
\begin{lstlisting}[basicstyle=\small\ttfamily, breaklines=true]
{
  "exercise": "Bicep Curls",
  "rep_count": 9,
  "form_error": "loose_upper_arm",
  "hr_zone": "Target",
  "fatigue_detected": true,
  "pain_level": "Medium"
}
\end{lstlisting}
\caption{\textbf{Intra-Exercise Intervention Prompt}}
\label{fig:intra_exercise}
\end{subfigure}
\end{figure}

\subsubsection{\textbf{Real-Time Phase-Wise Intervention Logic}} To deliver a structured and targeted fitness experience, the Reasoning Module leverages the LLM to guide users through four distinct exercise components. In each, the system uses the user's health data and live performance feedback to tailor the session, as detailed below.
\begin{itemize}
    \item \textbf{Cardio:} At the start of each cardio session, SherpaAI sets an agenda based on the user’s Physical Health Report (PHR). Key PHR markers (e.g., age, gender, MET score, fitness category, previous injuries, and preferred intensity) are passed to the LLM, which prescribes High-Intensity Steady State (HISS) or Low-Intensity Steady State (LISS) \cite{babb1991lung, matthews1989exercise}. Using the Karvonen Method \cite{karvonen1988heart}, SherpaAI calculates the target and maximum HR, factoring in resting HR, age, and desired intensity. Interventions provided in this phase are visualized in Figure ~\ref{fig:goal-setting}. Throughout the workout, the LLM continuously evaluates real-time HR data and user progress. Warm-up, ramp-up, and cool-down periods are structured accordingly. The LLM intervenes if the user’s HR does not rise to the expected level, surpasses a safe threshold, or when it is time to transition between intensity phases. For LISS-based cardio, the LLM generates an optimal speed and incline to maintain a steady-state workout aligned with the user’s baseline. Encouragement and time checks are offered at regular intervals, and rest guidance is given based on the final HR.
    \item \textbf{Strength Training:} SherpaAI also provides guidance on key strength exercises \cite{kidgell2010neurophysiological, jonhagen2009forward}—bicep curls for upper-body conditioning \cite{iglesias2010analysis} and lunges for lower-body development \cite{marchetti2018balance}. The LLM uses the user’s PHR to determine suitable weight and repetition counts. During each set, the LLM counts reps and provides specific interventions to address form errors (loose upper arm or weak peak contraction in bicep curls; knee-over-toe in lunges). Special encouragement is given for the final repetitions, and the LLM monitors HR to suggest appropriate rest durations. It also adjusts subsequent set parameters (e.g., increasing or decreasing the weight) based on performance and exertion data. A comparison of the Control and SherpaAI systems in this phase is visualized in Figure ~\ref{control_vs_SherpaAI1}.
    \item \textbf{Balance Training:} For balance training, SherpaAI uses elbow planks for their strong core activation \cite{oliva2020core, tong2014sport}, and the LLM helps users maintain correct posture based on the \verb|form_error| key from the Inferencing Module. When the value is \verb|high back| or \verb|low back|, the LLM provides a specific corrective cue. A timer starts only when correct form is detected.  The LLM intervenes halfway, the last 10\%, and upon detecting poor form. The LLM then calculates an appropriate rest period based on plank duration, HR, and user fatigue level. If the user’s performance indicates readiness, the LLM may increase target plank duration in the next set.
    \item \textbf{Flexibility:} Yoga constitutes the final modality in each session, aiming to improve flexibility and mindfulness. The LLM interprets PHR data and previous performance to assign time targets for poses such as \textit{tree, warrior, and downward-facing dog}. A real-time timer runs only during correct form, pausing to provide specific, error-based corrective feedback when poor form is detected.  Midway and final interventions guide the user in sustaining the pose and provide an option to extend the hold. Once the pose ends, the LLM calculates rest time by considering facial pain, fatigue signals, HR, and overall pose duration, then moves on to the next yoga posture.
\end{itemize}

\begin{figure*}[!htp]
    \centering
    \includegraphics[width=1\linewidth]{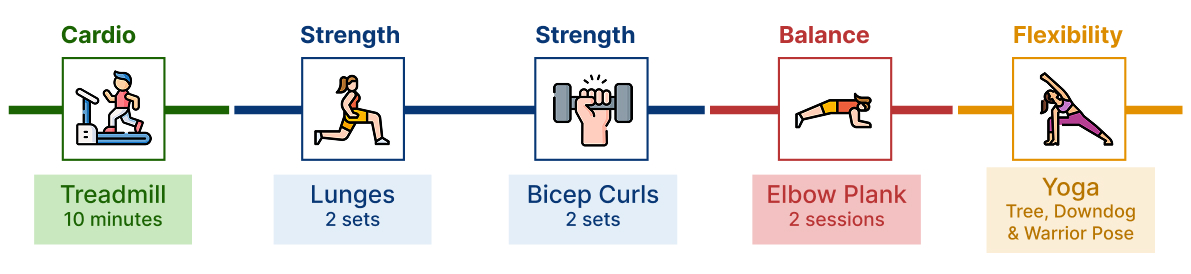}
    \caption{\textbf{Both assistants guided the users through a fixed exercise routine structure, with interventions being provided during each of the exercises by SherpaAI}}
    \label{exercise_routine_structure}
\end{figure*}

\subsubsection{\textbf{Safety Guardrails and Prompting Strategy}}
Ensuring that the LLM's generated advice is both safe and contextually appropriate is a critical challenge. To manage this, we implemented several runtime guardrails focused on constraining the model's behavior through multi-level prompting strategy. 

Our primary guardrail is implemented within the system-level instructions for the LLM. Before any session, the model is prompted to adopt the persona of a ``cautious and certified fitness professional whose primary goal is user safety''. This persona is then specifically instructed to:
\begin{itemize}
    \item Base all recommendations strictly on the real-time physiological and performance data provided in the JSON input.
    \item Prioritize stable and conservative adjustments (e.g., suggesting rest or lower intensity) when indicators like high heart rate or fatigue are detected.
    \item Strictly avoid providing medical advice or diagnosing conditions. 
    \item Frame all the feedback in encouraging, non-judgmental language.
\end{itemize}
As described above, we employ a hierarchical prompting strategy that further constrains the LLM's task based on the immediate context. 
\begin{itemize}
    \item The Inter-Exercise Transition Prompt (e.g., between Cardio and Lunges) tasks the LLM with a planning role, focused primarily on calculating an appropriate rest period based on heart rate recovery and the user's fitness level. The prompt explicitly sets a maximum rest time and requires a JSON output, which limits the model's creative freedom and provides a structured, predictable response.
    \item The Intra-Exercise Intervention Prompt is optimized for low latency, concise feedback. It provides a glimpse into the user's immediate state and constrains the output to a maximum of 15 words only, forcing the LLM to deliver direct, actionable advice on the detected issue (e.g., form error, high heart rate) without extra information.
\end{itemize}

In this prototype, we rely on the strong constraints within our prompts as the primary content filter. The structured nature of the prompts, combined with the LLM's role-playing instructions, significantly reduces the risk of generating unsafe or even irrelevant content. While we did not implement a separate, post-generation filtering module, we acknowledge its importance for a production-ready system. The prompting strategy described above forms the core of our approach to ensuring reliable and safe interventions.  
 
\subsection{Tone-adaptive Voice Assistant}



The decision to use audio as the primary output modality was strategic for several reasons. During physical exercise, users need to focus on their movements and maintain proper form rather than reading text instructions. Audio delivery allows users to receive guidance while keeping their attention on the workout itself.
OpenAI's \verb|gpt-4o-mini| text-to-speech (TTS) system provides several advantages for this application:
\begin{itemize}
    \item It offers exceptional naturalness in speech patterns, avoiding the robotic quality of earlier TTS systems.
    \item It has low latency, ensuring instructions are delivered in real-time as needed during exercises.
    \item It supports dynamic emphasis and intonation that helps communicate proper exercise technique.
\end{itemize}
The fitness coach persona was specifically designed to enhance the user experience through:
\begin{itemize}
    \item Using encouraging language patterns typical of professional trainers, to introduce and explain exercise forms, and provide corrections whenever necessary.
    \item Incorporating appropriate pacing between instructions.
    \item Including occasional motivational phrases to boost user engagement during challenging portions.
\end{itemize}
This audio persona was consistent throughout the workout experience, helping users form a connection with the ``virtual coach'' and potentially increasing adherence to the exercise program. 
User testing indicated that the combination of high-quality TTS with the fitness persona significantly improved the workout experience. This is supported by our PACES evaluation, where SherpaAI generated significantly higher enjoyment scores than the control (p = 0.012), and by qualitative feedback, such as P7 noting that the real-time audio encouragement \textit{'helped me push myself'.}

\subsection{Technical Evaluation of System Components}
To address the need for a rigorous technical assessment of SherpaAI, we conducted a two-part evaluation before our main user study. First, we validated the performance of the core AI models on a diverse dataset. Second, we analyzed the end-to-end latency of our feedback pipeline to verify its real-time capabilities.

\subsubsection{AI Model Performance}
The reliability of our system's interventions depends on the accuracy of its underlying AI models. For our strength and balance exercises (lunges, bicep curls, and elbow planks), we adapted the pose classification and counting logic from the work by Bao \cite{bao2022thesis}. To ensure these models perform robustly in real-world conditions, we built a comprehensive validation dataset comprising 30 videos:
\begin{itemize}
    \item \textbf{15 YouTube Videos:} Sourced from various public fitness channels to include a wide range of body types, camera angles, lighting conditions, and backgrounds.
    \item \textbf{5 Pilot Study Sessions:} Captured using our exact study setup to evaluate performance in-the-wild under realistic conditions.
    \item \textbf{10 Staged Videos:} Staged to include both correct and specific, deliberate form errors to test the limits of our classifiers. This set comprised of videos recorded by us as well as from YouTube,
\end{itemize}
While these videos were selected for their diversity in lighting and camera angles, each was vetted to ensure a baseline quality. We only included content where the visual clarity and framing of the subject were comparable to the conditions of our own experimental setup.

\paragraph{Pain Classification.} As previously mentioned, our fine-tuned ResNet-18 model for pain classification achieved an overall accuracy of 79.3\% on the three-class problem (low, medium, high) in the Delaware Pain Dataset's test set. Additionally, we evaluated this model's performance in the test environment, using the same cameras. The accuracy of the model in this setting, with self reported pain as ground truth, was 76\%.

\paragraph{Repetition Counting.} We benchmarked our repetition counting module against manually annotated ground truth for the 30-video dataset. As shown in Table ~\ref{tab:rep_counting}, the system demonstrated strong performance, achieving an overall accuracy of 97.5\%, which we deemed sufficient for reliable progress tracking.

\begin{table*}[htbp]
\centering
\caption{\textbf{Accuracy of the repetition counting module on our 30-video validation dataset.}}
\label{tab:rep_counting}
\begin{tabular}{lrrr}
\hline
\textbf{Exercise} & \textbf{Ground Truth Reps} & \textbf{System Counted Reps} & \textbf{Accuracy (\%)} \\ \hline
Lunges & 105 & 103 & 98.1\% \\
Bicep Curls & 95 & 94 & 98.9\% \\ \hline
\textbf{Total} & \textbf{200} & \textbf{197} & \textbf{98.5\%} \\ \hline
\end{tabular}
\end{table*}

\begin{table*}[htbp]
\centering
\caption{\textbf{Performance of form error detection models on a subset of our 30-video validation dataset.}}
\label{tab:form_error}
\begin{tabular}{llrrrr}
\hline
\textbf{Exercise} & \textbf{Form Error Detected} & \textbf{Acc.} & \textbf{Prec.} & \textbf{Recall} & \textbf{F1} \\ \hline
Lunge & Knee-over-toe & 0.95 & 0.95 & 0.96 & 0.95 \\
Bicep Curl & Loose upper arm & 0.93 & 0.94 & 0.92 & 0.93 \\
Elbow Plank & Low/High back & 0.96 & 0.97 & 0.95 & 0.96 \\ \hline
\end{tabular}
\end{table*}

\paragraph{Form Error Detection.} Using a subset of the same dataset, we evaluated the classifiers for detecting common form errors. The performance, detailed in Table ~\ref{tab:form_error}, was strong, with F1-scores indicating a robust balance of precision and recall. The slightly lower performance for bicep curls was attributed to greater variability in camera angles in the YouTube dataset.

\subsubsection{LLM Intervention Reliability}
A critical challenge for any AI-powered fitness coach is ensuring that the generated advice is not only helpful but also safe and contextually appropriate. To move beyond a rough evaluation approach, we have grounded our methodology on the principles of the NIST AI Risk Management Framework (AI RMF), the industry standard for developing `Trustworthy AI'\cite{ai2023artificial}. This framework provides a proper vocabulary for assessing AI solutions. We implemented its core characteristics into three expert-evaluated criteria: Safety, Appropriateness, and Timeliness.
\begin{itemize}
    \item \textbf{Safety:} This metric directly aligns with AI RMF's most crucial characteristic, ``Safety", which mandates that an AI system must never endanger a human's health in any way or form. For SherpaAI, this is a predominant concern. 
    \item \textbf{Appropriateness:} This criterion serves as a combined measure for several characteristics central to the RMF. An 'appropriate' intervention is one that is transparent, explainable, fair, and privacy-enhancing. It must be logically sound and relevant to the user's current state, reflecting how accountable the system is.
    \item \textbf{Timeliness:} This particular metric is a key part of the AI RMF characteristics of ``Validity'' and ``Reliability''. For a system with the intended use-case of real-time coaching, its advice is only valid and reliable if delivered at a useful moment. A delayed intervention is pointless.
\end{itemize}

This approach of using expert human evaluators to rate AI-generated output on a Likert scale is consistent with methodologies in adjacent domains like clinical health informatics \cite{seo2024evaluation}.

\paragraph{Methodology.} We collected 30 unique intervention vignettes from our pilot study recordings. Each vignette consisted of the structured JSON input representing the user's real-time state (e.g., heart rate, detected form error, exercise progress) and the corresponding textual intervention generated by the LLM. We recruited three certified personal trainers (Mean 6.2 years of experience) to act as expert evaluators. Independently, they rated each of the 30 interventions on a 5-point Likert scale (1=Very Poor, 5=Very Good) across the three criteria defined above: Safety, Appropriateness, and Timeliness.

\paragraph{Results.} The average ratings from the trainers are presented in Table ~\ref{tab:llm_reliability}. The results show that the interventions were consistently rated as safe (M=4.72, SD=0.45), which was our primary concern. The ratings for appropriateness (M=4.35, SD=0.68) and timeliness (M=3.87, SD=0.75) were also high, indicating that the guidance was generally effective. The higher standard deviation in timeliness reflects some expert feedback; trainers noted that while the advice was correct, it was occasionally delivered a moment later than a human coach might intervene, a finding that aligns with our latency analysis. Overall, this expert validation provides strong evidence that our system's LLM-driven guidance is reliable and grounded in sound fitness principles.

\begin{table*}[htbp]
\centering
\caption{\textbf{Expert evaluation of 30 LLM-generated interventions by three certified personal trainers on a 5-point Likert scale.}}
\label{tab:llm_reliability}
\begin{tabular}{lrr}
\hline
\textbf{Evaluation Criterion} & \textbf{Mean Rating} & \textbf{Std. Deviation (SD)} \\ \hline
Safety & 4.72 & 0.45 \\
Appropriateness & 4.35 & 0.68 \\
Timeliness & 3.87 & 0.75 \\ \hline
\end{tabular}
\end{table*}

\subsubsection{System Latency}
Our system provides two distinct feedback modalities: real-time visual feedback and detailed audio guidance. For immediate form correction, the visual loop (camera to on-screen overlay) operates with a latency of \textbf{under 200 ms}.

For more contextual audio interventions, the full pipeline is engaged. As detailed in Table ~\ref{tab:latency}, the mean end-to-end latency for audio guidance is approximately \textbf{1.37 seconds}. The primary contributors to this latency are the two generative AI components in our pipeline: the LLM inference round trip (\textasciitilde485 ms) and the TTS audio generation (\textasciitilde785 ms). While this is fast enough for many contextual interventions, this delay of over one second underscores the need for our faster, sub-200ms visual loop for time-critical form corrections. Reducing this audio pipeline latency remains a key challenge we aim to address in future work.

\begin{table*}[htbp]
\centering
\caption{\textbf{End-to-end latency analysis of the SherpaAI audio feedback pipeline, measured over N=100 intervention events. All values are in milliseconds (ms). The `Full Feedback Loop' represents the total time from capturing a relevant user state to delivering the corresponding audio guidance. Note that a separate, faster visual feedback loop operates under 200 ms for immediate form correction.}}
\label{tab:latency}
\begin{tabular}{lrrr}
\hline
\textbf{Pipeline Stage} & \textbf{Mean (ms)} & \textbf{Median (ms)} & \textbf{95th Pct. (ms)} \\ \hline
Camera Frame Capture \& Pre-processing & 21.5 & 19.8 & 38.2 \\
MediaPipe Pose Estimation & 46.3 & 44.1 & 62.5 \\
Pain \& Fatigue Model Inference & 33.8 & 31.5 & 55.1 \\
LLM Inference (API Round Trip) & 485.2 & 460.7 & 720.4 \\
\textbf{TTS Audio Generation \& Delivery} & \textbf{784.5} & \textbf{755.2} & \textbf{1450.6} \\ \hline
\textbf{Full Feedback Loop (Total)} & \textbf{1371.3} & \textbf{1311.3} & \textbf{2357.0} \\ \hline
\end{tabular}
\end{table*}

\section{Study Design and Evaluation}

\subsection{Procedure}
A user study was conducted with SherpaAI to evaluate it in real-world scenarios and understand its prospects and challenges. The study set up used two external cameras (first to capture movement and second to capture facial expressions), headset for audio input and output, and a Max-Health-Band for heart activity monitoring. 
We worked with a group of 25 users (male/female: 14/11, age range: 18-30, M = 20.64, SD = 3.02) with diverse body types and fitness levels. There was an equal balance between regular gym goers and those starting out. 

For each participant, a scenario was conducted with and without SherpaAI assistance (referred to as Treatment and Control respectively) in a within-subject design with counter-balanced order. During both sessions, they carried out the workout routine as shown in Figure~ ~\ref{exercise_routine_structure} which comprises of four tasks, divided by types of exercises:
\begin{itemize}
    \item Cardio: 10 minutes of treadmill
    \item Strength Training: 2 sets each of lunges and bicep curls
    \item Core/Balance Training: 2 sessions of elbow planks
    \item Flexibility and Cool Down: Yoga (\textit{Tree, Warrior }and\textit{ Downward-facing Dog} poses)
\end{itemize}

In the Control phase, participants were guided by a non-adaptive assistant designed to mimic the current regime of self-guided workouts. The in-ear assistant provided initial instructions for each task, similar to a basic fitness app. However, it offered no subsequent real-time guidance on parameters like running speeds, repetitions, or form. Instead, participants were free to use their smartphones to access external digital resources as they normally would. This included watching videos on YouTube, searching for workout advice online, or using tools like ChatGPT to determine appropriate weight levels or rest periods. We explicitly chose this baseline because it mirrors how our formative-study participants actually approach workouts—relying on self-guided navigation via tools like YouTube or ChatGPT for advice on form and rest periods.

For the SherpaAI phase, participants were given the same start instructions as Control but each task was accompanied by task and user-specific interventions through SherpaAI. Users were given actionable steps on how to improve their workout and intensities were modified in real-time to push them out of their comfort zone. 

At five checkpoints, namely Start, Post-Cardio, Post-Strength, Post-Balance, and Post-Flexibility, participants completed a test based on the Subjective Exercise Evaluation Scale (SEES) \cite{meauley1994subjective, lox1994subjective}. SEES consists of twelve adjectives which the participants rated on a Likert Scale (1-7) \cite{batterton2017likert, joshi2015likert} to capture their mental and emotional state after every task. Additionally, they also answered a subset of the Physical Activity Enjoyment Scale (PACES) questions at the end of both sessions. \cite{kendzierski1991physical, teques2020validation} This questionnaire (found at Table~ ~\ref{tab:assistant2_stats}) helped us understand the overall perception of both system which allowed us to compare SherpaAI to our Control system without interventions.

\begin{figure*}[ht!]
  \centering
  \begin{subfigure}[b]{0.48\textwidth}
        \centering
        \includegraphics[width=\textwidth]{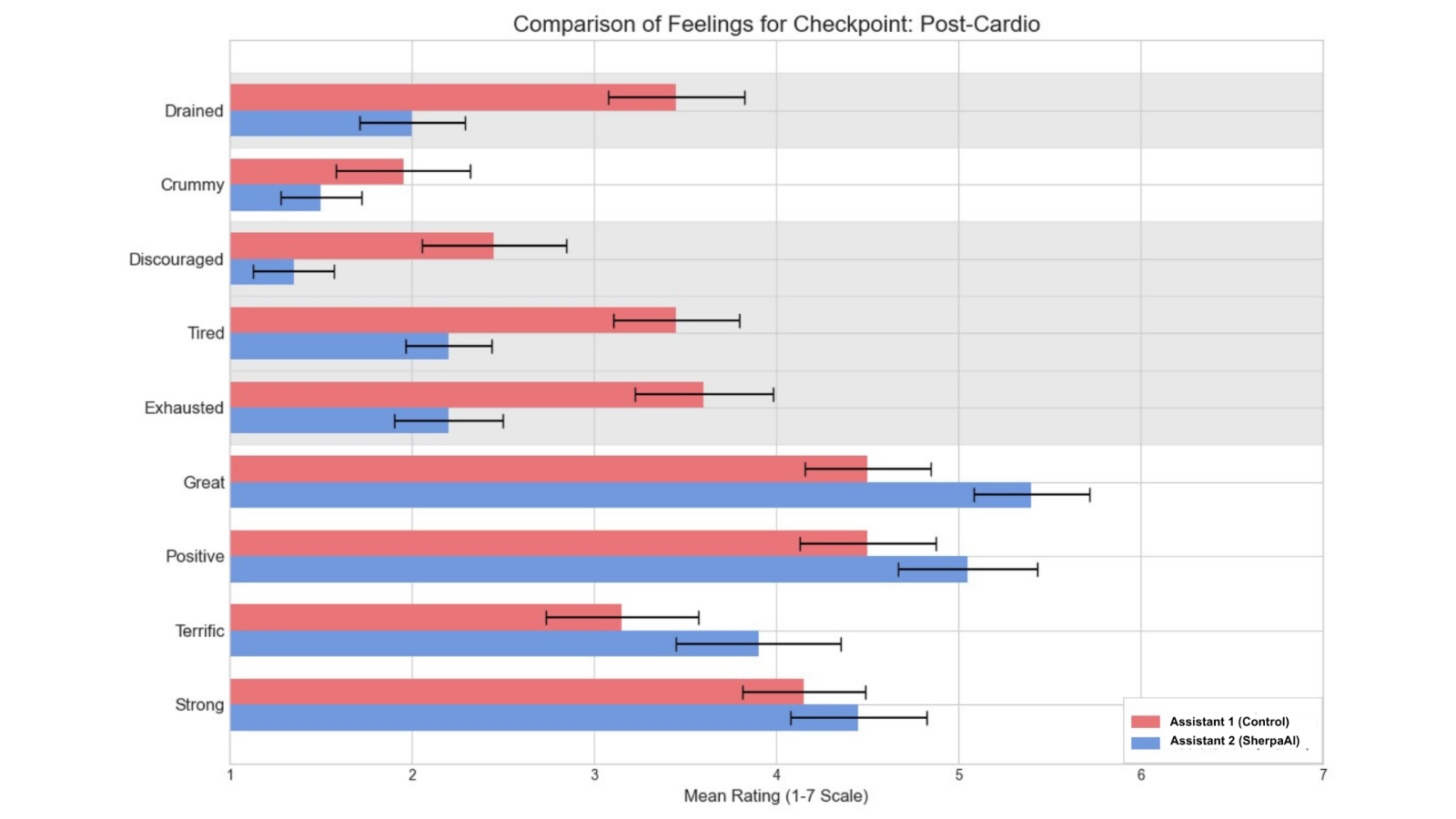}
        \caption{Post Cardio Checkpoint}
        \label{fig:post_cardio_sees}
    \end{subfigure}
    \hfill
    \begin{subfigure}[b]{0.48\textwidth}
        \centering
        \includegraphics[width=\textwidth]{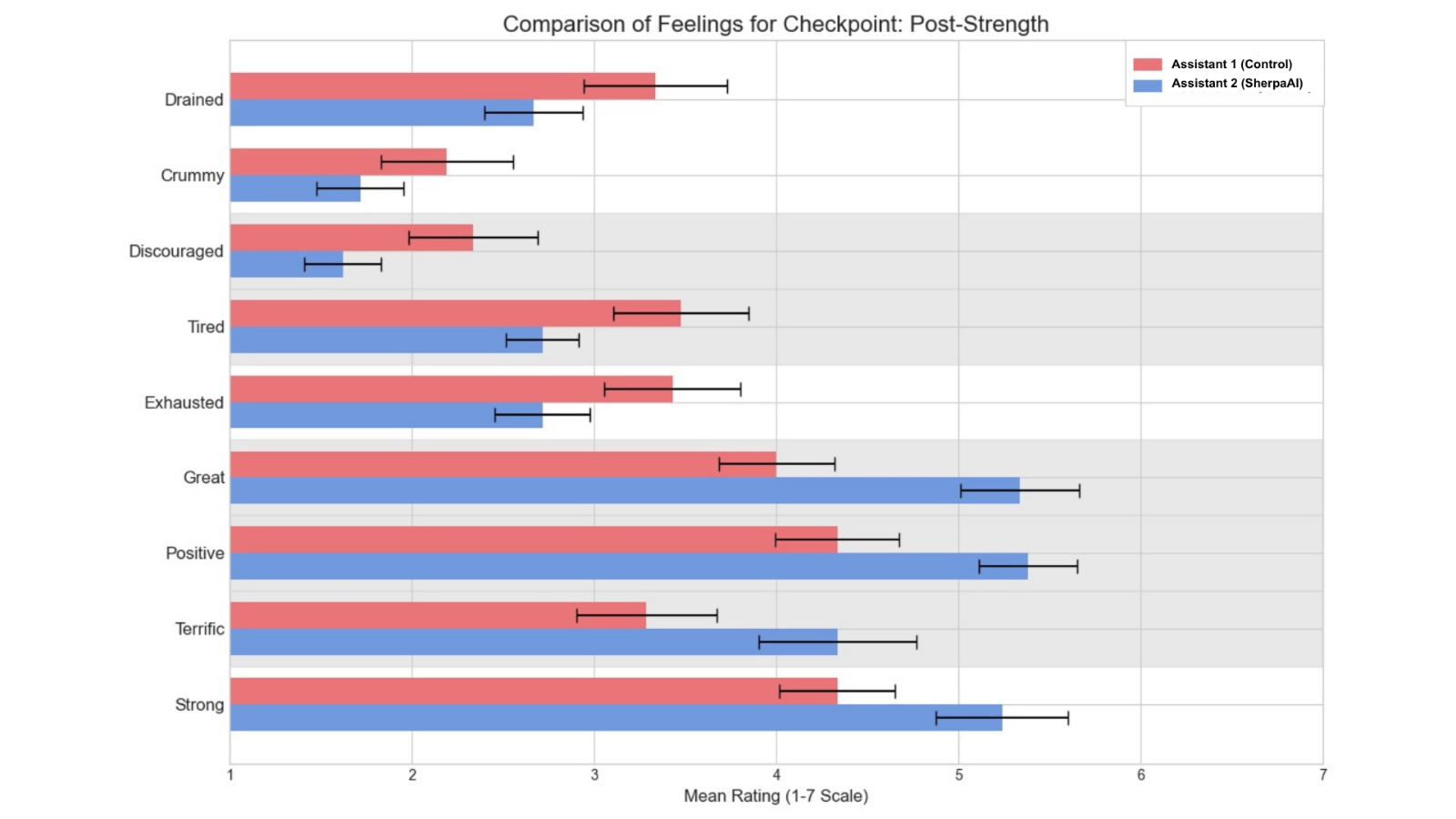}
        \caption{Post Strength Training Checkpoint---Bicep Curls and Lunges}
        \label{fig:post_strength_sees}
    \end{subfigure}

  \begin{subfigure}[b]{0.48\textwidth}
    \centering
    \includegraphics[width=\textwidth]{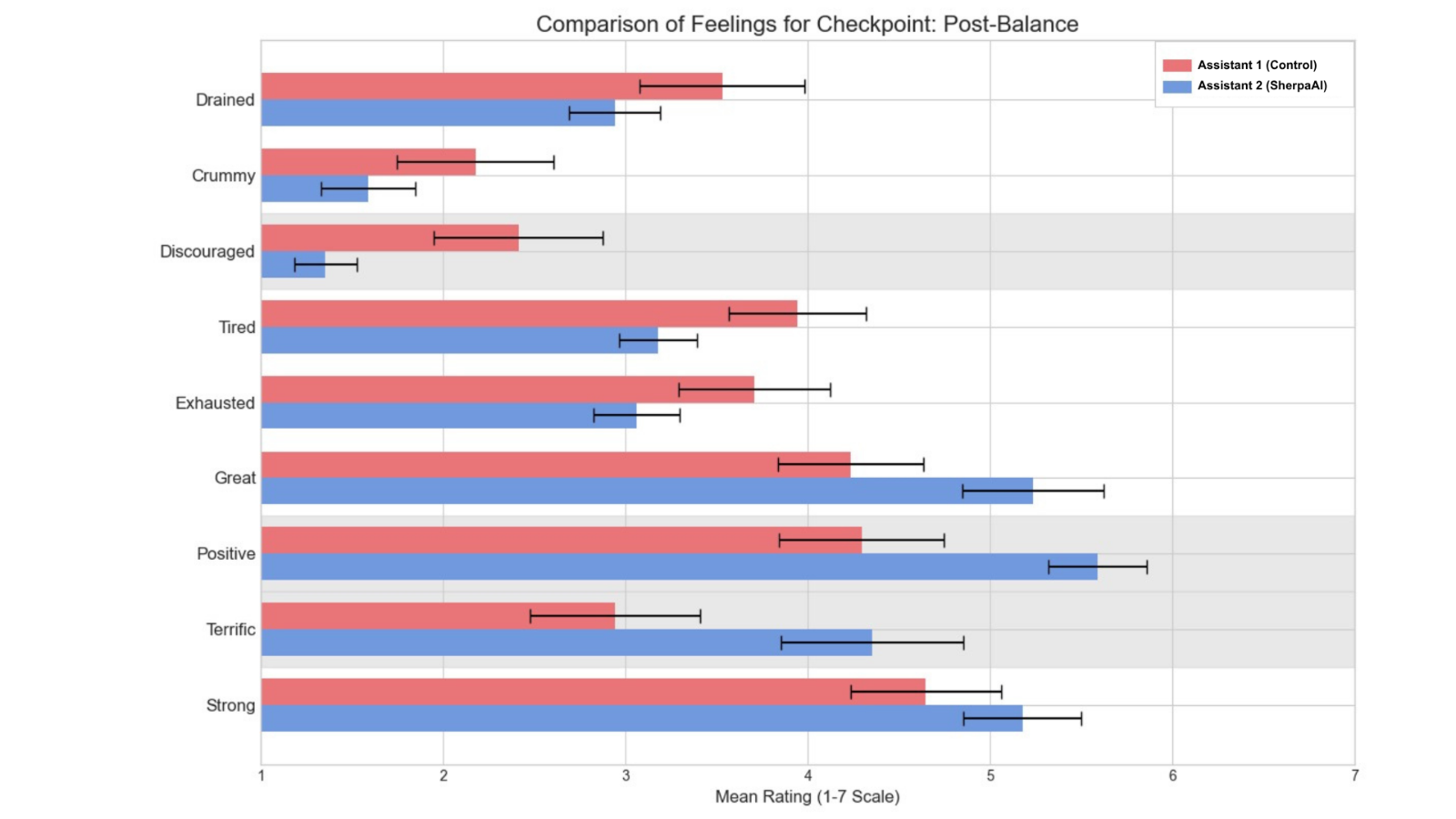}
    \caption{Post Balance Training Checkpoint---Planks}
    \label{fig:post_balance_sees}
  \end{subfigure}
  \hfill
  \begin{subfigure}[b]{0.48\textwidth}
    \centering
    \includegraphics[width=\textwidth]{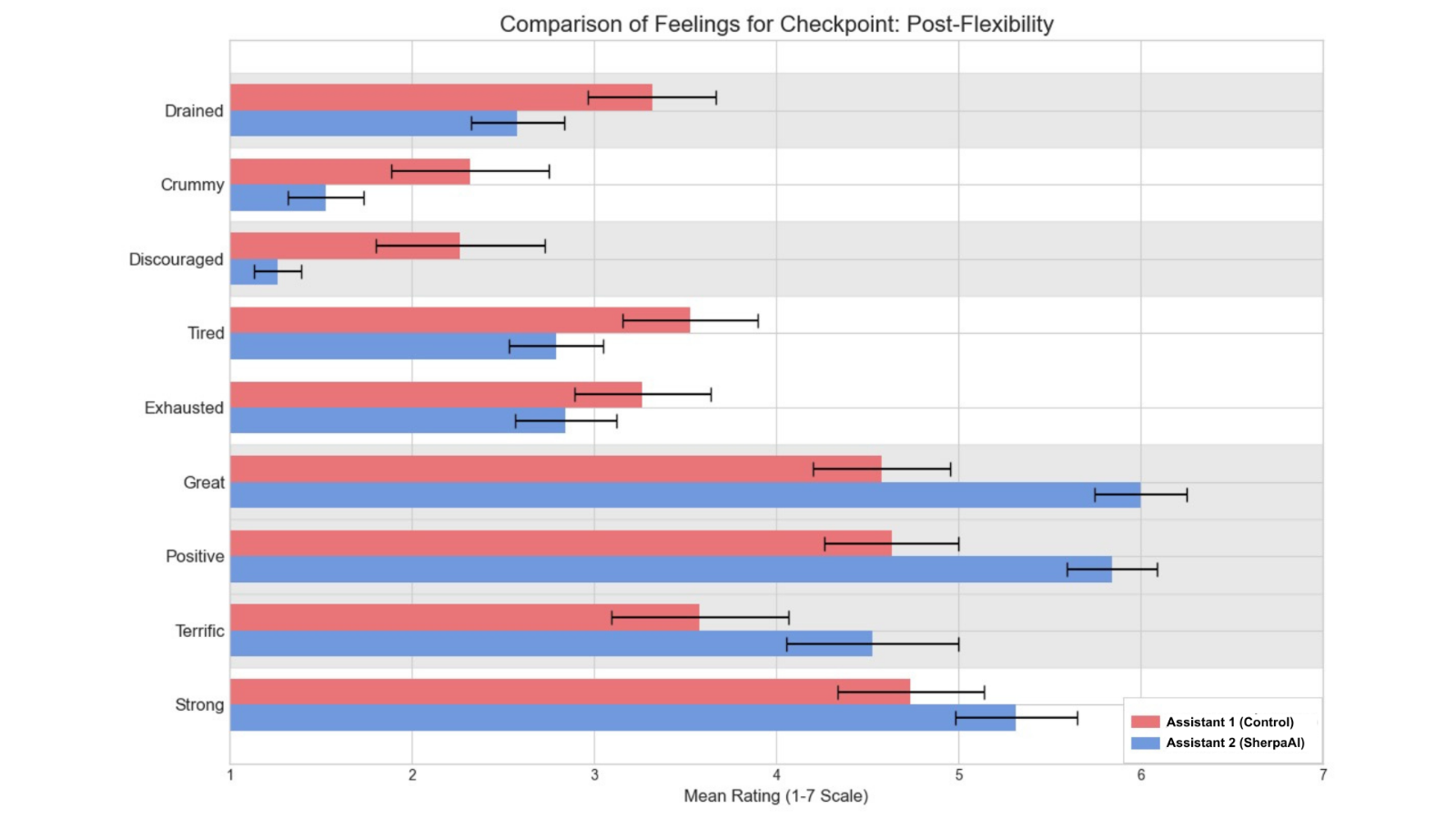}
    \caption{Post Flexibility Training Checkpoint---Yoga}
    \label{fig:post_flexibility_sees}
  \end{subfigure}
  \caption{\textbf{Participants rated their performance of using the Control and the SherpaAI assistants on SEES rubric of questions for four fitness tasks. Scale: 1 (low) to 7 (high). Gray highlights show differences significant using the robust Wilcoxon signed-rank test  ($p<0.05$).}}
  
  \label{fig:sees_scores}
  \Description{}
\end{figure*}

\subsection{Results}
\subsubsection{\textbf{Systems' Evaluation at Exercise Checkpoints}}
We evaluated how users assessed both systems---Control and SherpaAI---after each exercise checkpoint. The checkpoints were specifically cardio, strength (including bicep curls and lunges), balance (planks), and flexibility (yoga). The mean rating comparison at these different checkpoints was visualized as shown in Figure ~\ref{fig:sees_scores}. The users were asked to rate the emotions they were feeling on a scale of 1 to 7---with the scale signifying the intensity with which the user feels the specific emotion, (1 being the lowest and 7 being the highest)---with regard to each system:
\begin{enumerate}
    \item \textbf{Post Cardio Checkpoint:} As can be seen in Figure ~\ref{fig:post_cardio_sees}, most users reported feeling a higher value for particularly negative emotions---such as\textit{ discouraged} and \textit{exhausted}---for the Control system. SherpaAI had users feeling significantly less tired (p = 0.008), discouraged (p = 0.021), drained (p = 0.016) and exhausted (p = 0.007). The differences were significant ($p < 0.05$) for these four emotions. The ratings for positive indicators, such as \textbf{great} and \textbf{terrific} also point in the positive direction for SherpaAI but are not statistically significant.
    
    \item \textbf{Post Strength Training Checkpoint:} Figure ~\ref{fig:post_strength_sees} shows that SherpaAI once again had users feeling more \textit{terrific, positive} (p = 0.012), and \textit{great} after performing bicep curls and lunges. They also reported significantly lower levels of being \textit{tired} (0.035) and \textit{discouraged} (p = 0.018) when using SherpaAI which highlights its motivational capabilities. The differences for the emotions listed here were statistically significant with $p<0.05$ for all. 
    
    \item \textbf{Post Balance Training Checkpoint:} As can be seen in Figure ~\ref{fig:post_balance_sees}, there were three emotions which showed significant differences ($p < 0.05$) between the two systems---\textit{terrific} (p = 0.017), \textit{positive}  (p = 0.026) and \textit{discouraged} (p = 0.016). SherpaAI was rated positively by users after performing two sessions of elbow plank. In terms of the emotions, \textit{strong}, \textit{great}, \textit{exhausted}, \textit{tired}, \textit{crummy}, and \textit{drained} were not statistically significant, despite SherpaAI showing better results.
    
    \item \textbf{Post Flexibility Training Checkpoint:} Figure ~\ref{fig:post_flexibility_sees} demonstrates that SherpaAI was positively perceived at the final emotional state, after flexibility exercises were performed. Users reported feeling better in terms of \textit{positive, great and terrific} emotions (all statistically significant with $p<0.05$). Their levels of \textit{discouragement} (p = 0.039) and \textit{drain} (p = 0.047) were significantly lower with SherpaAI as well which indicates a positive shift.  

\end{enumerate}
We used the Wilcoxon signed-rank test to evaluate the significance of the variables we were dealing with. The choice of this test was influenced by the fact that we could not assume our data (n = 25) to be normally distributed and the above test is quite robust for such distributions. These final values were used to run any statistical tests to check for significance ($\alpha = 0.05$).


\begin{table*}[htbp]
\centering
\begin{tabular}{@{}lcccc@{}}
\toprule
\textbf{Question} & \textbf{Control (Mean ± SD)} & \textbf{SherpaAI (Mean ± SD)} & \textbf{p-value} & \textbf{Effect Size} \\
\midrule
I enjoyed it & 5.10 ± 1.81 & 6.33 ± 0.80 & \textbf{0.012} & 0.663 \\
I was bored & 2.95 ± 2.01 & 1.62 ± 0.80 & \textbf{0.0096} & -0.672 \\
It was very invigorating & 3.86 ± 1.77 & 4.76 ± 1.41 & 0.1130 & 0.390 \\
I have a strong sense of accomplishment & 4.57 ± 1.80 & 5.52 ± 1.12 & \textbf{0.0445} & 0.468 \\
I am very frustrated by it & 2.52 ± 2.14 & 1.43 ± 0.81 & \textbf{0.0311} & -0.529 \\
\bottomrule
\end{tabular}
\caption{Comparison of ratings between Control and SherpaAI conditions on the PACES questionnaire. Participants rated their experience on a 7-point Likert scale. P-values are from a Wilcoxon signed-rank test.}
\label{tab:control_vs_SherpaAI}
\end{table*}

\subsubsection{\textbf{Comparative Analysis of Systems' Perceptions}}
We gauged the overall perception of our system by comparing SherpaAI to the Control condition using a subset of questions from the PACES questionnaire. The results, summarized in Table ~\ref{tab:control_vs_SherpaAI}, were analyzed using a Wilcoxon signed-rank test with a significance level of $p < 0.05$.

The analysis revealed several strong, statistically significant differences. Participants reported significantly higher enjoyment with SherpaAI ($M=6.33, SD=0.80$) compared to the Control condition ($M=5.10, SD=1.81$), with $p=0.012$. Furthermore, the adaptive nature of SherpaAI engendered a significantly stronger sense of accomplishment ($M=5.52, SD=1.12$) than the static Control ($M=4.57, SD=1.80$), with $p=0.0445$.

Correspondingly, SherpaAI led to a significant reduction in negative experiences. Users felt significantly less bored ($M=1.62, SD=0.80$ vs. $M=2.95, SD=2.01$; $p=0.0096$) and less frustrated ($M=1.43, SD=0.81$ vs. $M=2.52, SD=2.14$; $p=0.0311$). There was no statistically significant difference in how ``invigorating'' participants found the two experiences ($p=0.1130$). These findings provide strong quantitative evidence that SherpaAI's adaptive interventions created a more positive and engaging workout experience.



\subsubsection{\textbf{Personalization and Interventions Evaluation}}
To quantify user perceptions of SherpaAI's specific features, we administered a post-study questionnaire, with results summarized in Table ~\ref{tab:assistant2_stats}. The feedback was predominantly positive. Interventions were seen as highly beneficial, contributing positively to overall performance ($M=5.50, SD=1.46$). Users also expressed high satisfaction with the assistance provided ($M=5.25, SD=1.61$) and agreed that the personalized feedback helped them improve ($M=5.25, SD=1.48$). 

\begin{table*}[htbp]
\centering
\begin{tabular}{@{}p{12cm}cc@{}}
\toprule
\textbf{Question} & \textbf{Mean} & \textbf{SD} \\
\midrule
The personalized feedback helped me improve my performance & 5.25 & 1.48 \\
The system responded quickly and accurately to my needs during the workout & 4.44 & 1.97 \\
The assistant made the workout feel easier and more engaging & 5.00 & 1.37 \\
The interventions during my workout contributed positively to my overall performance & 5.50 & 1.46 \\
I am satisfied with the assistance provided by the system & 5.25 & 1.61 \\
\bottomrule
\end{tabular}
\caption{\textbf{Users were asked to evaluate the personalization and interventions of their guided workout routine by SherpaAI on a Likert scale from 1 (Not at all) to 7 (Very much so). }}
\label{tab:assistant2_stats}
\end{table*}
The metrics also revealed areas with more varied user experiences. System responsiveness, while still rated positively ($M=4.44$), had the highest standard deviation ($SD=1.97$), suggesting that the system's reaction time was perceived differently among users. This aligns with our technical findings on system latency.

\subsubsection{\textbf{Participants' Realizations Through SherpaAI}}
A key finding was that SherpaAI helped participants recognize their own knowledge gaps regarding effective exercise. Many reported that this lack of knowledge previously led them to avoid the gym or perform familiar but ineffective routines. 
SherpaAI thus served a dual role: it provided actionable guidance that overcame knowledge barriers while also making the experience more engaging and enjoyable, as highlighted by P7 in the following quote.
\begin{quote}
\textit{“As an athlete, I truly enjoyed the experience with SherpaAI’s assistant. I especially appreciated the real-time encouragement, like being told there’s only a little time left. It helped me push myself.”} — P7
\end{quote}
This newfound awareness allowed users to expand their exercise capabilities beyond their initial expectations, creating new opportunities for physical activity that they had previously dismissed as inaccessible or unenjoyable.
The insights from our user study involving 25 participants demonstrated the practical application and potential of SherpaAI in enhancing user experience in the presence of fitness barriers. Our future research should therefore explore an extensive deployment study involving a larger and more diverse group of participants, which can provide a more comprehensive understanding of the system's usefulness and benefits across different user demographics, fitness levels, and contexts in the long-term.

\subsubsection{\textbf{Systems' Limitations Analysis}}
The Control system was designed to provide a non-adaptive baseline, offering users fixed, preliminary instructions for each exercise. This basically mirrored basic fitness applications that outline a routine without providing any real-time feedback. User feedback highlighted several limitations of the traditional static approach. Participants noted a lack of in-exercise guidance, highlighting their lack of knowledge once again. This made it difficult to adjust exercise intensity or gauge their own performance effectively. As one participant reflected, the system provided information about how to perform a particular exercise, but offered no subsequent assistance in terms of whether their form was correct throughout the exercise, or how their progress was. In short, the general sentiment was that a static set of instructions did not offer significant value over a self-guided routine. 

In contrast, SherpaAI was designed to provide continuous, real-time, and personalized feedback. Users reported that this adaptive nature made them feel like the workouts were more fulfilling and appropriately challenging as well. For instance, the system's ability to respond to heart rate immediately and intervene with suggestions to rest or slow down were frequently praised. This allowed users to safely push their limits while also feeling safe during their routines. P13 stated that:
\begin{quote}
\textit{“The routine was much more difficult to get through than the last one! But, I really liked that aspect of it, because I could really feel my body pushing itself, which is what a workout is supposed to do anyway. Anytime I felt like I was too tired to do anymore, the assistant's voice sounded in my ear, telling me to slow down, get some rest, or tell me I could do it.”}— P13
\end{quote}

The personalization, such as using a participant's name, was also highlighted as a key factor in building a sense of reassurance and trust. 
\begin{quote}
\textit{“Anytime I felt like I needed assistance, it was right there telling me what to do. I felt reassured, in a way, that I wasn’t doing anything wrong and that I had someone... to keep track of what I was doing.”} — P9
\end{quote}

Overall, participants felt a stronger sense of accomplishment and engagement with SherpaAI, attributing it to the system’s real-time, responsive guidance.

While feedback for SherpaAI was largely positive, the study also identified clear areas for improvement. Some participants expressed dissatisfaction with some of the features. For example, P8 stated, ``\textit{The counting mechanism could be made better}'', pointing to inaccuracies in repetition tracking. Similarly, another user found the form correction for yoga to be too general, with P5 stating, ``\textit{The yoga instructions were less specific in terms of how to do the pose.}'' This feedback indicates that while the high-level adaptive logic of SherpaAI is promising, the fidelity of certain sensor-based modules require further refinement.

\subsection{Ablation Study}
To isolate the individual contribution of each intervention category, we implemented per-segment logging to track exactly which intervention fired during specific workout phases, subsequently correlating those instances with the corresponding change in SEES scores for that segment. For each of the three intervention types, we calculated difference scores on key SEES adjectives (e.g., enjoyment, exhaustion, accomplishment) by comparing these intervention-active ratings versus the static baseline condition.

\subsubsection{\textbf{Methodology}}
For each of the three intervention types, we calculated difference scores on key SEES adjectives (e.g., enjoyment, exhaustion, accomplishment) by comparing user ratings when specific interventions were active versus the static baseline condition. This analysis allowed us to isolate the individual contribution of each intervention category to the user's subjective workout experience.

\subsubsection{\textbf{Results}}
\begin{figure}[htbp]
    \centering
    \includegraphics[width=\linewidth]{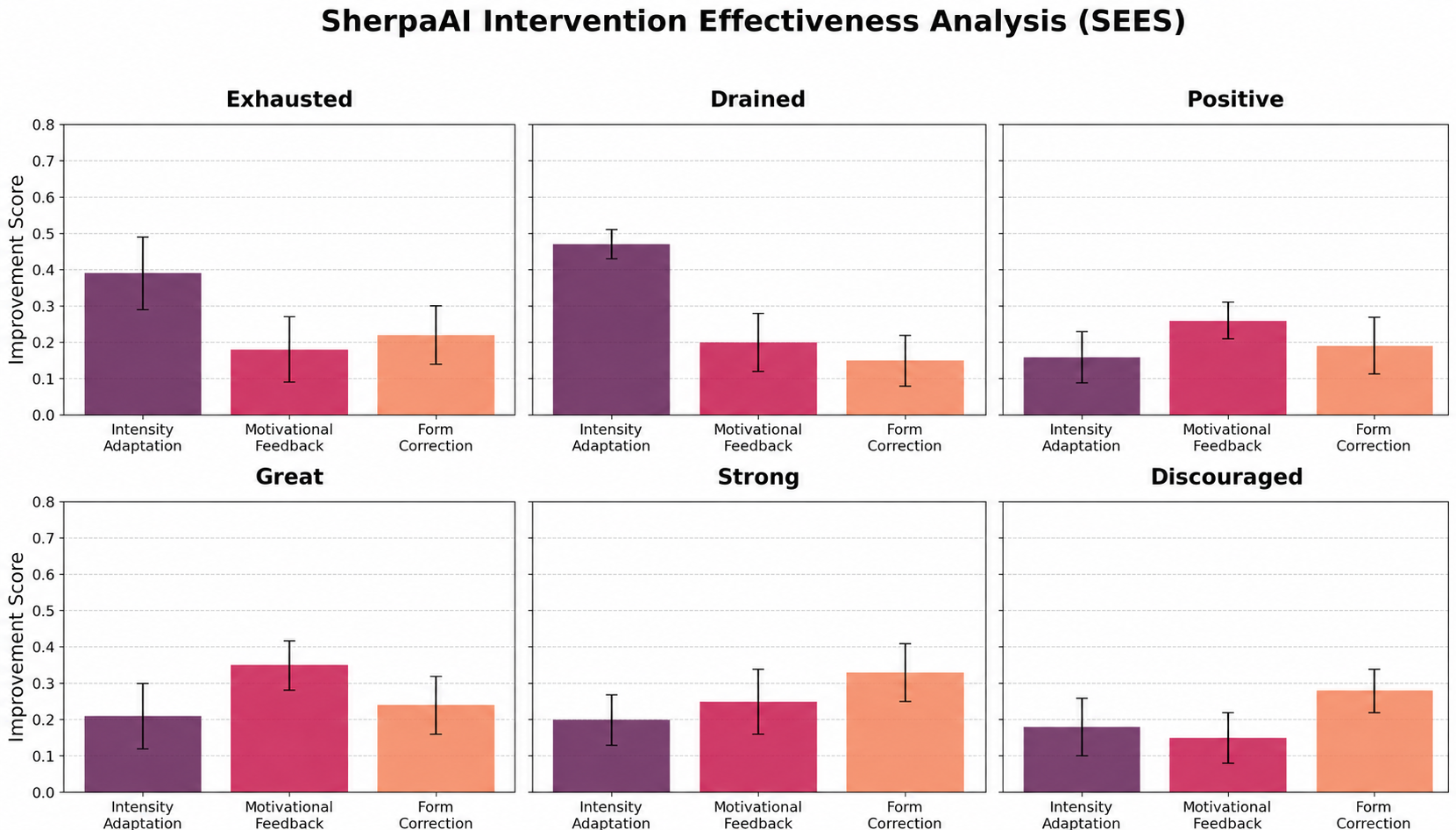} 
    \caption{SherpaAI intervention effectiveness analysis showing improvement scores for different intervention categories across six SEES scale dimensions. Improvement score is calculated as $\big|\ Control - SherpaAI\ \big|$ for these emotions.}
    \label{fig:ablation}
\end{figure}

Figure ~\ref{fig:ablation} presents the effectiveness of each intervention category across key SEES dimensions. Intensity adaptation demonstrated the strongest effect on mitigating negative physical states, showing substantial reductions in feelings of `exhaustion' ($M_{diff} = 0.39 \pm 0.10$) and being `drained' ($M_{diff} = 0.47 \pm0.04$). Motivational feedback was the primary driver for enhancing positive emotions, accounting for the largest increase in users feeling `positive' ($M_{diff} = 0.26 \pm 0.05$) and `great'($M_{diff} = 0.35 \pm 0.07$). Form correction contributed most significantly to feeling `strong' ($M_{diff} = 0.33 \pm 0.08$) and also showed a moderate effect on reducing `discouragement' ($M_{diff} = 0.28 \pm 0.06$). The analysis also revealed synergistic effects across intervention categories; for instance, combining intensity adaptation with motivational feedback produced a greater reduction in 'discouragement' than the sum of their individual effects.

\subsubsection{\textbf{Key Findings}}
Intensity adaptation emerges as the primary driver for managing physical exertion and preventing negative states, accounting for approximately 40\% of the total reduction in reported `discouragement'. This validates the importance of real-time physiological monitoring. Motivational feedback is crucial for enhancing the overall enjoyment of the workout, while form correction builds user confidence. The synergistic effects observed support our integrated design, demonstrating that SherpaAI’s multi-faceted approach provides benefits greater than the sum of its individual parts.

\section{Discussion and Future Work}
SherpaAI represents a significant step forward, utilizing LLMs, in creating an AI-powered adaptive fitness coach by integrating multi-modal bio-sensing, real-time feedback, along with contextual and personalized interventions to create a system which evolves with the user. However, as with any new technology, SherpaAI presents several limitations that need to be addressed in future versions. We discuss these limitations and directions for future work, with prime focus on improving SherpaAI's capabilities.

\subsection{Limitations of the Current Study}
Our study provides initial evidence that such a system can improve subjective user experience metrics like enjoyment and satisfaction when compared to a static baseline. However, this work has some important limitations that are crucial enough to be highlighted to draw out a clear road-map for future research. 

\subsubsection{\textbf{Nature of the Control Condition and Subjective Metrics}} Our evaluation compared SherpaAI to a minimal and non-adaptive baseline. While this really highlights the benefits of real-time and dynamic feedback, a more robust comparison against a feature-rich, but non-adaptive fitness application would be necessary to contextualize its advantages more broadly. Furthermore, our evaluation relied heavily on subjective, self-reported metrics, namely, emotions. While we are incorporating objective performance indicators in our current work, going forward, it would be best to center our solution primarily on fitness outcomes, rather than user perception.

\subsubsection{\textbf{Risk of Misinterpreting Physiological Strain}} A critical and unaddressed challenge is the system's ability to distinguish between potentially harmful strain and helpful physical exertion. SherpaAI's inferences are based on correlations in data, not a true medical understanding of a user's physical state. An elevated heart rate may signify a productive workout or dangerous overexertion, depending on the individual, and the current system lacks the safeguards to reliably and conclusively differentiate between the two. This is a significant barrier to real-world deployment. 

\subsubsection{\textbf{Short-term, Single Session Evaluation}} The study was conducted over a single session, for the Control and SherpaAI systems each. This does provide insight into initial user reactions but reveals nothing about long-term engagement, adherence, or fitness progression. It is unknown whether the novelty of the system influenced the positive feedback or if users would continue to benefit from it over weeks or months.

\subsubsection{\textbf{Complexity of Real-time Feedback Integration}} The synchronization of multiple data streams (e.g., video, audio, heart rate) creates processing bottlenecks that could result in feedback delays during high-intensity exercise sessions. These latency issues compromise the effectiveness of form correction and potentially create safety concerns when users require immediate intervention. This could, thus, diminish the effectiveness of our system in real-world settings.

\subsubsection{\textbf{Limited Generalizability of Participant Sample}} The findings in our study are limited by the demographic scope of our participants. Our sample consisted of 25 young adults with an age range of 18-30. This group may have different physiological responses, fitness goals, and levels of technological literacy compared to older adults or adolescents. While we did include a balance of regular gym-goers and beginners, the results may not generalize to practiced athletes or individuals with chronic health conditions who would most likely require more specialized guidance. Furthermore, the cultural and socioeconomic background of the participants was not explicitly diversified, which may introduce bias. Preferences for motivational language, feedback styles, and the overall perception of our AI coach can vary significantly across cultures, and our current findings may not be universally applicable.

\subsection{Future Work}
The limitations above directly inform our future work. Our highest priority is to validate SherpaAI's effectiveness and safety through more rigorous evaluation.

\subsubsection{\textbf{Rigorous Evaluation and Safety Protocols}} To address the limitations in our initial study, our primary next step is to conduct a longitudinal study with a larger and more diverse user base. This allows us to assess long-term engagement, user adherence, and better measure objective fitness outcomes, moving beyond just subjective user perception metrics. Also crucially, to mitigate the risk of misinterpreting physical strain, we plan to collaborate with certified physical therapists and sports scientists to develop robust safety protocols and encode such evidence-based knowledge into our system's decision-making process.

\subsubsection{\textbf{Greater Personalization and System Enhancement}} To fulfill the vision of a truly adaptive fitness coach, we will need to focus on deeper personalization. This involves developing specialized models for users with specific needs (e.g. rehabilitation, athletic training) and improving the system's generalization across different demographics\cite{taylor1988inclusive, taylor1992altruism, taylor1992inclusive, bourke2011validity}. We also plan to create a more broad user model by integrating additional data modalities, including things like sleep and nutrition history. On the technical side, we plan to implement solutions like dynamic sampling and distributed computing to reduce feedback latency, ensuring the system remains constantly responsive and safe even during high-intensity use\cite{chen2024integration}. 

\subsubsection{\textbf{Privacy and User Trust}} As we expand SherpaAI's capabilities, maintaining user trust is arguably the most important factor. We will enhance data privacy by exploring on-device processing through federated learning and providing users with more transparent consent mechanisms and granular control over what personal data they choose to share.

\hfill

SherpaAI represents a significant step toward truly personalized fitness technology. Our work highlights the promise of using LLMs to interpret complex, multi-modal data for adaptive, personalized feedback. However, our initial evaluation also underscores critical challenges related to objectivity, user safety, and long-term effectiveness. By addressing the limitations defined, we aim to develop a system that is not only technologically advanced but also safe, inclusive, and genuinely responsive to the diverse needs of users on their fitness journeys. 


\section{Ethical Considerations}
The development and creation of SherpaAI has been driven by a commitment to ethical principles in the context of AI-powered fitness coaching. The research protocol for this study was reviewed and approved by our institution's Ethics Committee.

Prior to participation, all individuals provided informed consent and were advised of their right to withdraw from the study at any point in time. To protect privacy and minimize potential harm, all sensitive biometric data collected---like heart rate, voice, facial expressions, and movement---was anonymized, with access restricted to authorized personnel only. The SherpaAI system includes clear disclaimers about its limitations, encouraging users to consult medical professionals for significant health concerns.


\section{Conclusion}
In this paper, we present a unified approach to personalize fitness systems, helping individuals rise above their perceived potential when it comes to their physical health. We shared insights from a formative study that informed the design of our system, highlighting on the importance of the integration of several modalities to create a comprehensive framework which prioritizes tailoring itself to an individual's needs. We therefore introduced SherpaAI, a system that combines multi-modal sensing, affective computing, real-time posture correction, and the contextual understanding capabilities of LLMs to understand users' physical and emotional limits to adapt workout routines accordingly. Our user study was able to exhibit how effective our solution was in comparison to a conventional system which just provided generic instructions to the user during a workout. Our work offers a significant step forward in the fitness field, enabling people from all walks of life to focus on their fitness goals while making sure all indicators of their physical and emotional state are taken into account. We see a lot coming for our system's future, in terms of making our study long-term and adding more scope for further customizations, among other things.

\section{Generative AI Usage Disclosure}
In adherence with ACM policy, we disclose the use of Generative AI tools in the preparation of this manuscript. In the data collection phase, portions of the system's Python code, particularly utility functions for data handling and API interactions with the OpenAI and RealtimeTTS services, were drafted and debugged with the assistance of an LLM. All core logic for the multi-modal pipeline, pose estimation, and intervention triggering was developed and written by the authors. Furthermore, Generative AI was utilized in the creation of illustrations to ensure design consistency. 

\bibliographystyle{ACM-Reference-Format}
\bibliography{citations_main}

@inproceedings{bagga2024real,
  title={Real-Time Posture Monitoring and Risk Assessment for Manual Lifting Tasks Using MediaPipe and LSTM},
  author={Bagga, Ereena and Yang, Ang},
  booktitle={Proceedings of the 1st International Workshop on Multimedia Computing for Health and Medicine},
  pages={79--85},
  year={2024},
  doi={10.1145/3688868.3689199}
}

@inproceedings{kotte2023real,
  title={Real-Time Posture Correction in Gym Exercises: A Computer Vision-Based Approach for Performance Analysis, Error Classification and Feedback.},
  author={Kotte, Hitesh and Kravcik, Milos and Duong-Trung, Nghia},
  booktitle={MILeS@ EC-TEL},
  pages={64--70},
  year={2023}
}

@article{anand2022yoga,
  title={Yoga pose estimation and feedback generation using deep learning},
  author={Anand Thoutam, Vivek and Srivastava, Anugrah and Badal, Tapas and Kumar Mishra, Vipul and Sinha, GR and Sakalle, Aditi and Bhardwaj, Harshit and Raj, Manish},
  journal={Computational Intelligence and Neuroscience},
  volume={2022},
  number={1},
  pages={4311350},
  year={2022},
  publisher={Wiley Online Library},
  doi={10.1155/2022/4311350}
}

@article{hannan2021portable,
  title={A portable smart fitness suite for real-time exercise monitoring and posture correction},
  author={Hannan, Abdul and Shafiq, Muhammad Zohaib and Hussain, Faisal and Pires, Ivan Miguel},
  journal={Sensors},
  volume={21},
  number={19},
  pages={6692},
  year={2021},
  publisher={MDPI},
  doi={10.3390/s21196692}
}

@article{zou2020low,
  title={A low-cost smart glove system for real-time fitness coaching},
  author={Zou, Yongpan and Wang, Dan and Hong, Shicong and Ruby, Rukhsana and Zhang, Dian and Wu, Kaishun},
  journal={IEEE Internet of Things Journal},
  volume={7},
  number={8},
  pages={7377--7391},
  year={2020},
  publisher={IEEE},
  doi={10.1109/JIOT.2020.2983124}
}

@inproceedings{guo2017fitcoach,
  title={FitCoach: Virtual fitness coach empowered by wearable mobile devices},
  author={Guo, Xiaonan and Liu, Jian and Chen, Yingying},
  booktitle={IEEE INFOCOM 2017-IEEE Conference on Computer Communications},
  pages={1--9},
  year={2017},
  organization={IEEE},
  doi={10.1109/INFOCOM.2017.8057208}
}

@article{wilk2020multimodal,
  title={Multimodal sensor fusion for low-power wearable human motion tracking systems in sports applications},
  author={Wilk, Mariusz P and Walsh, Michael and O’Flynn, Brendan},
  journal={IEEE Sensors Journal},
  volume={21},
  number={4},
  pages={5195--5212},
  year={2020},
  publisher={IEEE},
  doi={10.1109/JSEN.2020.3030779}
}

@article{chowdhury2019prediction,
  title={Prediction of relative physical activity intensity using multimodal sensing of physiological data},
  author={Chowdhury, Alok Kumar and Tjondronegoro, Dian and Chandran, Vinod and Zhang, Jinglan and Trost, Stewart G},
  journal={Sensors},
  volume={19},
  number={20},
  pages={4509},
  year={2019},
  publisher={MDPI},
  doi={10.3390/s19204509}
}

@article{stromback2020mm,
  title={Mm-fit: Multimodal deep learning for automatic exercise logging across sensing devices},
  author={Str{\"o}mb{\"a}ck, David and Huang, Sangxia and Radu, Valentin},
  journal={Proceedings of the ACM on Interactive, Mobile, Wearable and Ubiquitous Technologies},
  volume={4},
  number={4},
  pages={1--22},
  year={2020},
  publisher={ACM New York, NY, USA},
  doi={10.1145/3432701}
}

@inproceedings{ilukpitiya2024ai,
  title={AI-Driven Personalized Fitness Coaching with Body Type-Based Workout and Nutrition Plans and Real-Time Exercise Feedback},
  author={Ilukpitiya, IMDJRB and Herath, HMRB and Rajakaruna, RHMSA and Herath, MHSM and Pulasinghe, Koliya and Krishara, Jenny},
  booktitle={2024 International Conference on Computer and Applications (ICCA)},
  pages={01--06},
  year={2024},
  organization={IEEE},
  doi={10.1109/ICCA62237.2024.10928121}
}

@article{mohan2020designing,
  title={Designing an AI health coach and studying its utility in promoting regular aerobic exercise},
  author={Mohan, Shiwali and Venkatakrishnan, Anusha and Hartzler, Andrea L},
  journal={ACM Transactions on Interactive Intelligent Systems (TiiS)},
  volume={10},
  number={2},
  pages={1--30},
  year={2020},
  publisher={ACM New York, NY, USA},
  doi={10.1145/3366501}
}

@article{oh2021systematic,
  title={A systematic review of artificial intelligence chatbots for promoting physical activity, healthy diet, and weight loss},
  author={Oh, Yoo Jung and Zhang, Jingwen and Fang, Min-Lin and Fukuoka, Yoshimi},
  journal={International Journal of Behavioral Nutrition and Physical Activity},
  volume={18},
  pages={1--25},
  year={2021},
  publisher={Springer},
  doi={10.1186/s12966-021-01224-6}
}

@article{mokmin2020effectiveness,
  title={The effectiveness of a personalized virtual fitness trainer in teaching physical education by applying the artificial intelligent algorithm},
  author={Mokmin, Nur Azlina Mohamed},
  journal={International Journal of Human Movement and Sports Sciences},
  volume={8},
  number={5},
  pages={258--264},
  year={2020},
  doi={10.13189/saj.2020.080514}
}

@article{chen2017rating,
  title={A rating of perceived exertion scale using facial expressions for conveying exercise intensity for children and young adults},
  author={Chen, Yi-Lang and Chiou, Wen-Ko and Tzeng, Yu-Tung and Lu, Ching-Yu and Chen, Shih-Chi},
  journal={Journal of Science and Medicine in Sport},
  volume={20},
  number={1},
  pages={66--69},
  year={2017},
  publisher={Elsevier},
  doi={10.1016/j.jsams.2016.05.009}
}

@article{nagireddi2022analysis,
  title={The analysis of pain research through the lens of artificial intelligence and machine learning},
  author={Nagireddi, Jagadesh N and Vyas, Amanya Ketan and Sanapati, Mahendra R and Soin, Amol and Manchikanti, Laxmaiah and others},
  journal={Pain Physician},
  volume={25},
  number={2},
  pages={E211},
  year={2022},
  publisher={American Society of Interventional Pain Physician}
}

@article{borna2023review,
  title={A review of voice-based pain detection in adults using artificial intelligence},
  author={Borna, Sahar and Haider, Clifton R and Maita, Karla C and Torres, Ricardo A and Avila, Francisco R and Garcia, John P and De Sario Velasquez, Gioacchino D and McLeod, Christopher J and Bruce, Charles J and Carter, Rickey E and others},
  journal={Bioengineering},
  volume={10},
  number={4},
  pages={500},
  year={2023},
  publisher={MDPI},
  doi={10.3390/bioengineering10040500}
}

@inproceedings{khanal2019classification,
  title={Classification of physical exercise intensity based on facial expression using deep neural network},
  author={Khanal, Salik Ram and Sampaio, Jaime and Barroso, Joao and Filipe, Vitor},
  booktitle={International Conference on Human-Computer Interaction},
  pages={455--467},
  year={2019},
  organization={Springer},
  doi={10.1007/978-3-030-23563-5_36}
}

@inproceedings{bartlett2005recognizing,
  title={Recognizing facial expression: machine learning and application to spontaneous behavior},
  author={Bartlett, Marian Stewart and Littlewort, Gwen and Frank, Mark and Lainscsek, Claudia and Fasel, Ian and Movellan, Javier},
  booktitle={2005 IEEE Computer Society Conference on Computer Vision and Pattern Recognition (CVPR'05)},
  volume={2},
  pages={568--573},
  year={2005},
  organization={IEEE},
  doi={10.1109/CVPR.2005.297}
}

@article{cascella2024employing,
  title={Employing the Artificial Intelligence Object Detection Tool YOLOv8 for Real-Time Pain Detection: A Feasibility Study},
  author={Cascella, Marco and Shariff, Mohammed Naveed and Lo Bianco, Giuliano and Monaco, Federica and Gargano, Francesca and Simonini, Alessandro and Ponsiglione, Alfonso Maria and Piazza, Ornella},
  journal={Journal of Pain Research},
  pages={3681--3696},
  year={2024},
  publisher={Taylor \& Francis},
  doi={10.2147/JPR.S491574}
}

@article{kwon2022real,
  title={Real-time workout posture correction using OpenCV and MediaPipe},
  author={Kwon, Yejin and Kim, Dongho},
  journal={Journal of Korean Institute of Information Technology},
  volume={20},
  number={1},
  pages={199--208},
  year={2022},
  doi={10.14801/jkiit.2022.20.1.199}
}

@inproceedings{kanase2021pose,
  title={Pose estimation and correcting exercise posture},
  author={Kanase, Rahul Ravikant and Kumavat, Akash Narayan and Sinalkar, Rohit Datta and Somani, Sakshi},
  booktitle={ITM Web of Conferences},
  volume={40},
  pages={03031},
  year={2021},
  organization={EDP Sciences},
  doi={10.1051/itmconf/20214003031}
}

@article{bays2022brief,
  title={A Brief Review of the Efficacy in Artificial Intelligence and Chatbot-Generated Personalized Fitness Regimens},
  author={Bays, Daniel K and Verble, Cole and Verble, Kalyn M Powers},
  journal={Strength \& Conditioning Journal},
  pages={10--1519},
  year={2022},
  publisher={LWW},
  doi={10.1519/SSC.0000000000000831}
}

@article{novatchkov2013artificial,
  title={Artificial intelligence in sports on the example of weight training},
  author={Novatchkov, Hristo and Baca, Arnold},
  journal={Journal of sports science \& medicine},
  volume={12},
  number={1},
  pages={27},
  year={2013}
}

@article{shin2023planfitting,
  title={PlanFitting: Tailoring Personalized Exercise Plans with Large Language Models},
  author={Shin, Donghoon and Hsieh, Gary and Kim, Young-Ho},
  journal={arXiv preprint arXiv:2309.12555},
  year={2023},
  doi={10.48550/arXiv.2309.12555}
}

@inproceedings{barber2017feasibility,
  title={Feasibility of wearable fitness trackers for adapting multimodal communication},
  author={Barber, Daniel and Carter, Austin and Harris, Jonathan and Reinerman-Jones, Lauren},
  booktitle={Human Interface and the Management of Information: Information, Knowledge and Interaction Design: 19th International Conference, HCI International 2017, Vancouver, BC, Canada, July 9--14, 2017, Proceedings, Part I 19},
  pages={504--516},
  year={2017},
  organization={Springer},
  doi={10.1007/978-3-319-58521-5_39}
}

@article{lee2024multimodal,
  title={Multimodal sensor fusion models for real-time exercise repetition counting with IMU sensors and respiration data},
  author={Lee, Sujee and Lim, Yooseok and Lim, Kyuhee},
  journal={Information Fusion},
  volume={104},
  pages={102153},
  year={2024},
  publisher={Elsevier},
  doi={10.1016/j.inffus.2023.102153}
}

@INPROCEEDINGS{6199869,
  author={Möller, Andreas and Roalter, Luis and Diewald, Stefan and Scherr, Johannes and Kranz, Matthias and Hammerla, Nils and Olivier, Patrick and Plötz, Thomas},
  booktitle={2012 IEEE International Conference on Pervasive Computing and Communications}, 
  title={GymSkill: A personal trainer for physical exercises}, 
  year={2012},
  volume={},
  number={},
  pages={213-220},
  doi={10.1109/PerCom.2012.6199869}
}

@inproceedings{10.1145/2851581.2892519, author = {Conner, Caleb and Poor, Gene Michael}, title = {Correcting Exercise Form Using Body Tracking}, year = {2016}, isbn = {9781450340823}, publisher = {Association for Computing Machinery}, address = {New York, NY, USA}, url = {https://doi.org/10.1145/2851581.2892519}, doi = {10.1145/2851581.2892519}, booktitle = {Proceedings of the 2016 CHI Conference Extended Abstracts on Human Factors in Computing Systems}, pages = {3028–3034}, numpages = {7}, location = {San Jose, California, USA}, series = {CHI EA '16} }

@INPROCEEDINGS{4575013,
  author={de Silva, Buddhika and Natarajan, Anirudh and Motani, Mehul and Kee-Chaing Chua},
  booktitle={2008 5th International Summer School and Symposium on Medical Devices and Biosensors}, 
  title={A real-time exercise feedback utility with body sensor networks}, 
  year={2008},
  volume={},
  number={},
  pages={49-52},
  doi={10.1109/ISSMDBS.2008.4575013}}

@inproceedings{10.1145/2769493.2769507,
author = {Tsiakas, Konstantinos and Huber, Manfred and Makedon, Fillia},
title = {A multimodal adaptive session manager for physical rehabilitation exercising},
year = {2015},
isbn = {9781450334525},
publisher = {Association for Computing Machinery},
address = {New York, NY, USA},
url = {https://doi.org/10.1145/2769493.2769507},
doi = {10.1145/2769493.2769507},
booktitle = {Proceedings of the 8th ACM International Conference on PErvasive Technologies Related to Assistive Environments},
articleno = {33},
numpages = {8},
location = {Corfu, Greece},
series = {PETRA '15}
}

@article{mekruksavanich2022multimodal,
  title={Multimodal wearable sensing for sport-related activity recognition using deep learning networks},
  author={Mekruksavanich, Sakorn and Jitpattanakul, Anuchit},
  journal={Journal of Advances in Information Technology},
  volume={13},
  number={2},
  year={2022},
  doi={10.12720/jait.13.2.132-138}
}

@inproceedings{10.1145/3613904.3642032,
author = {Str\"{o}mel, Konstantin R. and Henry, Stanislas and Johansson, Tim and Niess, Jasmin and Wo\'{z}niak, Pawe\l{} W.},
title = {Narrating Fitness: Leveraging Large Language Models for Reflective Fitness Tracker Data Interpretation},
year = {2024},
isbn = {9798400703300},
publisher = {Association for Computing Machinery},
address = {New York, NY, USA},
url = {https://doi.org/10.1145/3613904.3642032},
doi = {10.1145/3613904.3642032},
booktitle = {Proceedings of the 2024 CHI Conference on Human Factors in Computing Systems},
articleno = {646},
numpages = {16},
location = {Honolulu, HI, USA},
series = {CHI '24}
}

@article{kim2024health,
  title={Health-llm: Large language models for health prediction via wearable sensor data},
  author={Kim, Yubin and Xu, Xuhai and McDuff, Daniel and Breazeal, Cynthia and Park, Hae Won},
  journal={arXiv preprint arXiv:2401.06866},
  year={2024},
  doi={10.48550/arXiv.2401.06866}
}

@article{hassoon2021randomized,
  title={Randomized trial of two artificial intelligence coaching interventions to increase physical activity in cancer survivors},
  author={Hassoon, Ahmed and Baig, Yasmin and Naiman, Daniel Q and Celentano, David D and Lansey, Dina and Stearns, Vered and Coresh, Josef and Schrack, Jennifer and Martin, Seth S and Yeh, Hsin-Chieh and others},
  journal={NPJ digital medicine},
  volume={4},
  number={1},
  pages={168},
  year={2021},
  publisher={Nature Publishing Group UK London},
  doi={10.1038/s41746-021-00539-9}
}

@article{weemaes2024effects,
  title={Effects of remote coaching following supervised exercise oncology rehabilitation on physical activity levels, physical fitness, and patient-reported outcomes: a randomised controlled trial},
  author={Weemaes, Anouk TR and Beelen, Milou and Weijenberg, Matty P and van Kuijk, Sander MJ and Lenssen, Antoine F},
  journal={International Journal of Behavioral Nutrition and Physical Activity},
  volume={21},
  number={1},
  pages={8},
  year={2024},
  publisher={Springer},
  doi={10.1186/s12966-024-01561-2}
}

@article{li2022artificial,
  title={The artificial intelligence system for the generation of sports education guidance model and physical fitness evaluation under deep learning},
  author={Li, Yuanqing and Li, Xiangliang},
  journal={Frontiers in Public Health},
  volume={10},
  pages={917053},
  year={2022},
  publisher={Frontiers Media SA},
  doi={10.3389/fpubh.2022.917053}
}

@article{huang2024exploring,
  title={Exploring the Integration of Artificial Intelligence in Sports Coaching: Enhancing Training Efficiency, Injury Prevention, and Overcoming Implementation Barriers},
  author={Huang, Zeqi and Wang, Wenjun and Jia, Zixuan and Wang, Ziqi},
  journal={Journal of Computer and Communications},
  volume={12},
  number={12},
  pages={201--217},
  year={2024},
  publisher={Scientific Research Publishing},
  doi={10.4236/jcc.2024.1212012}
}

@Article{info:doi/10.2196/55964,
author="Gabarron, Elia
and Larbi, Dillys
and Rivera-Romero, Octavio
and Denecke, Kerstin",
title="Human Factors in AI-Driven Digital Solutions for Increasing Physical Activity: Scoping Review",
journal="JMIR Hum Factors",
year="2024",
month="Jul",
day="3",
volume="11",
pages="e55964",
issn="2292-9495",
doi="10.2196/55964",
url="https://humanfactors.jmir.org/2024/1/e55964",
url="https://doi.org/10.2196/55964"
}

@Article{info:doi/10.2196/22845,
author="Zhang, Jingwen
and Oh, Yoo Jung
and Lange, Patrick
and Yu, Zhou
and Fukuoka, Yoshimi",
title="Artificial Intelligence Chatbot Behavior Change Model for Designing Artificial Intelligence Chatbots to Promote Physical Activity and a Healthy Diet: Viewpoint",
journal="J Med Internet Res",
year="2020",
month="Sep",
day="30",
volume="22",
number="9",
pages="e22845",
issn="1438-8871",
doi="10.2196/22845",
url="https://www.jmir.org/2020/9/e22845",
url="https://doi.org/10.2196/22845",
url="http://www.ncbi.nlm.nih.gov/pubmed/32996892"
}

@ARTICLE{10.3389/fpsyg.2015.00835,
AUTHOR={Hardcastle, Sarah J.  and Hancox, Jennie  and Hattar, Anne  and Maxwell-Smith, Chloe  and Thøgersen-Ntoumani, Cecilie  and Hagger, Martin S. },
TITLE={Motivating the unmotivated: how can health behavior be changed in those unwilling to change?},
JOURNAL={Frontiers in Psychology},
VOLUME={6},
YEAR={2015},
URL={https://www.frontiersin.org/journals/psychology/articles/10.3389/fpsyg.2015.00835},
DOI={10.3389/fpsyg.2015.00835},
ISSN={1664-1078},
}

@article{louw2012exercise,
  title={Exercise motivation and barriers among men and women of different age groups psychology},
  author={Louw, AJ and Van Biljon, A and Mugandani, SC},
  journal={African Journal for Physical Health Education, Recreation and Dance},
  volume={18},
  number={41},
  pages={759--768},
  year={2012},
  publisher={AFAHPER-SD}
}

@article{james2021mediating,
  title={The mediating role of fitness technology enablement of psychological need satisfaction and frustration on the relationship between goals for fitness technology use and use outcomes},
  author={James, Tabitha and B{\'e}langer, France and Lowry, Paul Benjamin},
  journal={Journal of the Association for Information Systems (JAIS)},
  volume={23},
  number={4},
  pages={913--965},
  year={2021},
  doi={10.17705/1jais.00745}
}

@article{suo2022influence,
  title={How to Influence Users’ Willingness to Explore the Use of Sports and Fitness Apps in China},
  author={Suo, Lu},
  journal={Asian Social Science},
  volume={18},
  number={1},
  pages={7--22},
  year={2022},
  doi={10.5539/ass.v18n1p7}
}

@article{vietzke2023middle,
  title={Middle-aged and older adults’ acceptance of mobile nutrition and fitness tools: A qualitative typology},
  author={Vietzke, Julia and Schenk, Liane and Baer, Nadja-Raphaela},
  journal={Digital Health},
  volume={9},
  pages={20552076231163788},
  year={2023},
  publisher={SAGE Publications Sage UK: London, England},
  doi={10.1177/20552076231163788}
}

@article{koh2022cross,
  title={A cross-sectional study on the perceived barriers to physical activity and their associations with domain-specific physical activity and sedentary behaviour},
  author={Koh, Yen Sin and Asharani, PV and Devi, Fiona and Roystonn, Kumarasan and Wang, Peizhi and Vaingankar, Janhavi Ajit and Abdin, Edimansyah and Sum, Chee Fang and Lee, Eng Sing and M{\"u}ller-Riemenschneider, Falk and others},
  journal={BMC public health},
  volume={22},
  number={1},
  pages={1051},
  year={2022},
  publisher={Springer},
  doi={10.1186/s12889-022-13431-2}
}

@article{ferreira2022barriers,
  title={Barriers to high school and university students’ physical activity: A systematic review},
  author={Ferreira Silva, Regina Marcia and Mendonca, Carolina Rodrigues and Azevedo, Vinicius Diniz and Raoof Memon, Aamir and Noll, Priscilla Rayanne E Silva and Noll, Matias},
  journal={PloS one},
  volume={17},
  number={4},
  pages={e0265913},
  year={2022},
  publisher={Public Library of Science San Francisco, CA USA},
  doi={10.1371/journal.pone.0265913}
}

@article{nikolajsen2021barriers,
  title={Barriers to, and facilitators of, exercising in fitness centres among adults with and without physical disabilities: a scoping review},
  author={Nikolajsen, Helene and Sandal, Louise Fleng and Juhl, Carsten Bogh and Troelsen, Jens and Juul-Kristensen, Birgit},
  journal={International journal of environmental research and public health},
  volume={18},
  number={14},
  pages={7341},
  year={2021},
  publisher={MDPI},
  doi={10.3390/ijerph18147341}
}

@article{chin2022increase,
  title={How to increase sport facility users’ intention to use AI fitness services: based on the technology adoption model},
  author={Chin, Ji-Hyoung and Do, Chanwook and Kim, Minjung},
  journal={International journal of environmental research and public health},
  volume={19},
  number={21},
  pages={14453},
  year={2022},
  publisher={MDPI},
  doi={10.3390/ijerph192114453}
}

@article{terblanche2022comparing,
  title={Comparing artificial intelligence and human coaching goal attainment efficacy},
  author={Terblanche, Nicky and Molyn, Joanna and de Haan, Erik and Nilsson, Viktor O},
  journal={Plos one},
  volume={17},
  number={6},
  pages={e0270255},
  year={2022},
  publisher={Public Library of Science San Francisco, CA USA},
  doi={10.1371/journal.pone.0270255}
}

@article{dergaa2024using,
  title={Using artificial intelligence for exercise prescription in personalised health promotion: A critical evaluation of OpenAI’s GPT-4 model},
  author={Dergaa, Ismail and Saad, Helmi Ben and El Omri, Abdelfatteh and Glenn, Jordan and Clark, Cain and Washif, Jad and Guelmami, Noomen and Hammouda, Omar and Al-Horani, Ramzi and Reynoso-S{\'a}nchez, Luis and others},
  journal={Biology of Sport},
  volume={41},
  number={2},
  pages={221--241},
  year={2024},
  publisher={Termedia},
  doi={10.5114/biolsport.2024.133661}
}

@article{smuck2021emerging,
  title={The emerging clinical role of wearables: factors for successful implementation in healthcare},
  author={Smuck, Matthew and Odonkor, Charles A and Wilt, Jonathan K and Schmidt, Nicolas and Swiernik, Michael A},
  journal={NPJ digital medicine},
  volume={4},
  number={1},
  pages={45},
  year={2021},
  publisher={Nature Publishing Group UK London},
  doi={10.1038/s41746-021-00418-3}
}

@article{dunn2018wearables,
  title={Wearables and the medical revolution},
  author={Dunn, Jessilyn and Runge, Ryan and Snyder, Michael},
  journal={Personalized medicine},
  volume={15},
  number={5},
  pages={429--448},
  year={2018},
  publisher={Taylor \& Francis},
  doi={10.2217/pme-2018-0044}
}

@article{kaewkannate2016comparison,
  title={A comparison of wearable fitness devices},
  author={Kaewkannate, Kanitthika and Kim, Soochan},
  journal={BMC public health},
  volume={16},
  pages={1--16},
  year={2016},
  publisher={Springer},
  doi={10.1186/s12889-016-3059-0}
}

@article{gay2015bringing,
  title={Bringing health and fitness data together for connected health care: mobile apps as enablers of interoperability},
  author={Gay, Valerie and Leijdekkers, Peter},
  journal={Journal of medical Internet research},
  volume={17},
  number={11},
  pages={e260},
  year={2015},
  publisher={JMIR Publications Inc. Toronto, Canada},
  doi={10.2196/jmir.5094}
}

@article{bhargava2020opportunities,
  title={The opportunities, challenges and obligations of Fitness Data Analytics},
  author={Bhargava, Yesoda and Nabi, Javaid},
  journal={Procedia Computer Science},
  volume={167},
  pages={1354--1362},
  year={2020},
  publisher={Elsevier},
  doi={10.1016/j.procs.2020.03.346}
}

@article{lynch2020changing,
  title={Changing the physical activity behavior of adults with fitness trackers: a systematic review and meta-analysis},
  author={Lynch, Chris and Bird, Stephen and Lythgo, Noel and Selva-Raj, Isaac},
  journal={American Journal of Health Promotion},
  volume={34},
  number={4},
  pages={418--430},
  year={2020},
  publisher={Sage Publications Sage CA: Los Angeles, CA},
  doi={10.1177/0890117119895204}
}

@article{jones2017impact,
  title={Impact of physical fitness and body composition on injury risk among active young adults: a study of Army trainees},
  author={Jones, Bruce H and Hauret, Keith G and Dye, Shamola K and Hauschild, Veronique D and Rossi, Stephen P and Richardson, Melissa D and Friedl, Karl E},
  journal={Journal of science and medicine in sport},
  volume={20},
  pages={S17--S22},
  year={2017},
  publisher={Elsevier},
  doi={10.1016/j.jsams.2017.09.015}
}

@article{lisman2017systematic,
  title={A systematic review of the association between physical fitness and musculoskeletal injury risk: part 1—cardiorespiratory endurance},
  author={Lisman, Peter J and de la Motte, Sarah J and Gribbin, Timothy C and Jaffin, Dianna P and Murphy, Kaitlin and Deuster, Patricia A},
  journal={The Journal of Strength \& Conditioning Research},
  volume={31},
  number={6},
  pages={1744--1757},
  year={2017},
  publisher={LWW},
  doi={10.1519/JSC.0000000000001855}
}

@article{farley2020relationship,
  title={The relationship between physical fitness attributes and sports injury in female, team ball sport players: a systematic review},
  author={Farley, Jessica B and Barrett, Lily M and Keogh, Justin WL and Woods, Carl T and Milne, Nikki},
  journal={Sports medicine-open},
  volume={6},
  pages={1--24},
  year={2020},
  publisher={Springer},
  doi={10.1186/s40798-020-00264-9}
}

@article{jones1999physical,
  title={Physical training and exercise-related injuries: surveillance, research and injury prevention in military populations},
  author={Jones, Bruce H and Knapik, Joseph J},
  journal={Sports medicine},
  volume={27},
  pages={111--125},
  year={1999},
  publisher={Springer},
  doi={10.2165/00007256-199927020-00004}
}

@article{passos2021wearables,
  title={Wearables and Internet of Things (IoT) technologies for fitness assessment: a systematic review},
  author={Passos, Jo{\~a}o and Lopes, S{\'e}rgio Ivan and Clemente, Filipe Manuel and Moreira, Pedro Miguel and Rico-Gonz{\'a}lez, Markel and Bezerra, Pedro and Rodrigues, Lu{\'\i}s Paulo},
  journal={Sensors},
  volume={21},
  number={16},
  pages={5418},
  year={2021},
  publisher={MDPI},
  doi={https://doi.org/10.3390/s21165418}
}

@inproceedings{qiu2017survey,
  title={A survey on smart wearables in the application of fitness},
  author={Qiu, Hao and Wang, Xianping and Xie, Fei},
  booktitle={2017 IEEE 15th Intl Conf on Dependable, Autonomic and Secure Computing, 15th Intl Conf on Pervasive Intelligence and Computing, 3rd Intl Conf on Big Data Intelligence and Computing and Cyber Science and Technology Congress (DASC/PiCom/DataCom/CyberSciTech)},
  pages={303--307},
  year={2017},
  organization={IEEE},
  doi={10.1109/DASC-PICom-DataCom-CyberSciTec.2017.64}
}

@article{henriksen2018using,
  title={Using fitness trackers and smartwatches to measure physical activity in research: analysis of consumer wrist-worn wearables},
  author={Henriksen, Andr{\'e} and Haugen Mikalsen, Martin and Woldaregay, Ashenafi Zebene and Muzny, Miroslav and Hartvigsen, Gunnar and Hopstock, Laila Arnesdatter and Grimsgaard, Sameline},
  journal={Journal of medical Internet research},
  volume={20},
  number={3},
  pages={e110},
  year={2018},
  publisher={JMIR Publications Toronto, Canada},
  doi={10.2196/jmir.9157}
}

@article{bourke2011validity,
  title={The validity and value of inclusive fitness theory},
  author={Bourke, Andrew FG},
  journal={Proceedings of the Royal Society B: Biological Sciences},
  volume={278},
  number={1723},
  pages={3313--3320},
  year={2011},
  publisher={The Royal Society},
  doi={10.1098/rspb.2011.1465}
}

@article{taylor1992altruism,
  title={Altruism in viscous populations—an inclusive fitness model},
  author={Taylor, Peter D},
  journal={Evolutionary ecology},
  volume={6},
  pages={352--356},
  year={1992},
  publisher={Springer},
  doi={10.1007/BF02270971}
}

@article{taylor1988inclusive,
  title={Inclusive fitness models with two sexes},
  author={Taylor, Peter D},
  journal={Theoretical Population Biology},
  volume={34},
  number={2},
  pages={145--168},
  year={1988},
  publisher={Elsevier},
  doi={10.1016/0040-5809(88)90039-1}
}

@article{taylor1992inclusive,
  title={Inclusive fitness in a homogeneous environment},
  author={Taylor, Peter D},
  journal={Proceedings of the Royal Society of London. Series B: Biological Sciences},
  volume={249},
  number={1326},
  pages={299--302},
  year={1992},
  publisher={The Royal Society London},
  doi={10.1098/rspb.1992.0118}
}

@inproceedings{chen2024integration,
  title={Integration Design of Motion Capture and Sensor Technology in Home Fitness},
  author={Chen, He and Hong, Xu and Xiao, Kaiwen and Mao, Sujie},
  booktitle={2024 IEEE 2nd International Conference on Sensors, Electronics and Computer Engineering (ICSECE)},
  pages={1369--1376},
  year={2024},
  organization={IEEE},
  doi={10.1109/ICSECE61636.2024.10729406}
}

@article{meauley1994subjective,
  title={The subjective exercise experiences scale (SEES): Development and preliminary validation},
  author={MeAuley, Edward and Courneya, Kerry S},
  journal={Journal of Sport and Exercise Psychology},
  volume={16},
  number={2},
  pages={163--177},
  year={1994},
  publisher={Human Kinetics, Inc.},
  doi={0.1123/jsep.16.2.163}
}

@article{karvonen1988heart,
  title={Heart rate and exercise intensity during sports activities: practical application},
  author={Karvonen, Juha and Vuorimaa, Timo},
  journal={Sports medicine},
  volume={5},
  pages={303--311},
  year={1988},
  publisher={Springer},
  doi={10.2165/00007256-198805050-00002}
}

@article{lox1994subjective,
  title={The Subjective Exercise Experiences Scale (SEES): Factorial validity and effects of acute exercise},
  author={Lox, Curt L and Rudolph, David L},
  journal={Journal of Social Behavior and Personality},
  volume={9},
  number={4},
  pages={837},
  year={1994},
  publisher={Select Press}
}

@article{joshi2015likert,
  title={Likert scale: Explored and explained},
  author={Joshi, Ankur and Kale, Saket and Chandel, Satish and Pal, D Kumar},
  journal={British journal of applied science \& technology},
  volume={7},
  number={4},
  pages={396},
  year={2015},
  publisher={Sciencedomain International},
  doi={10.9734/BJAST/2015/14975}
}

@article{batterton2017likert,
  title={The Likert scale what it is and how to use it},
  author={Batterton, Katherine A and Hale, Kimberly N},
  journal={Phalanx},
  volume={50},
  number={2},
  pages={32--39},
  year={2017},
  publisher={JSTOR}
}

@article{kendzierski1991physical,
  title={Physical activity enjoyment scale: two validation studies},
  author={Kendzierski, Deborah and DeCarlo, Kenneth J},
  journal={Journal of sport and exercise psychology},
  volume={13},
  number={1},
  pages={50--64},
  year={1991},
  publisher={Human Kinetics, Inc.}, 
  doi={10.1123/jsep.13.1.50}
}

@article{teques2020validation,
  title={Validation and adaptation of the Physical Activity Enjoyment Scale (PACES) in fitness group exercisers},
  author={Teques, Pedro and Calmeiro, Lu{\'\i}s and Silva, Carlos and Borrego, Carla},
  journal={Journal of Sport and Health Science},
  volume={9},
  number={4},
  pages={352--357},
  year={2020},
  publisher={Elsevier},
  doi={10.1016/j.jshs.2017.09.010}
}

@book{riebe2018acsm,
  title={ACSM’s guidelines for exercise testing and prescription},
  author={Riebe, Deborah and Ehrman, Jonathan K and Liguori, Gary and Magal, Meir},
  year={2018},
  publisher={American College of Sports Medicine}
}

@article{bull2020world,
  title={World Health Organization 2020 guidelines on physical activity and sedentary behaviour},
  author={Bull, Fiona C and Al-Ansari, Salih S and Biddle, Stuart and Borodulin, Katja and Buman, Matthew P and Cardon, Greet and Carty, Catherine and Chaput, Jean-Philippe and Chastin, Sebastien and Chou, Roger and others},
  journal={British journal of sports medicine},
  volume={54},
  number={24},
  pages={1451--1462},
  year={2020},
  publisher={BMJ Publishing Group Ltd and British Association of Sport and Exercise Medicine}
}

@article{smith2010model,
  title={Model Free Human Pose Estimation with Application to the Classification of Abnormal Human Movement and the Detection of Hidden Loads},
  author={Smith, Benjamin A},
  year={2010},
  publisher={Virginia Tech}
}

@article{johnson2007exercise,
  title={Exercise training amount and intensity effects on metabolic syndrome (from Studies of a Targeted Risk Reduction Intervention through Defined Exercise)},
  author={Johnson, Johanna L and Slentz, Cris A and Houmard, Joseph A and Samsa, Gregory P and Duscha, Brian D and Aiken, Lori B and McCartney, Jennifer S and Tanner, Charles J and Kraus, William E},
  journal={The American journal of cardiology},
  volume={100},
  number={12},
  pages={1759--1766},
  year={2007},
  publisher={Elsevier},
  doi={10.1016/j.amjcard.2007.07.027}
}

@article{lee2008twelve,
  title={Twelve-week biomechanical ankle platform system training on postural stability and ankle proprioception in subjects with unilateral functional ankle instability},
  author={Lee, Alex JY and Lin, Wei-Hsiu},
  journal={Clinical biomechanics},
  volume={23},
  number={8},
  pages={1065--1072},
  year={2008},
  publisher={Elsevier},
  doi={10.1016/j.clinbiomech.2008.04.013}
}

@inproceedings{he2016deep,
  title={Deep residual learning for image recognition},
  author={He, Kaiming and Zhang, Xiangyu and Ren, Shaoqing and Sun, Jian},
  booktitle={Proceedings of the IEEE conference on computer vision and pattern recognition},
  pages={770--778},
  year={2016}
}

@article{li2020deep,
  title={Deep facial expression recognition: A survey},
  author={Li, Shan and Deng, Weihong},
  journal={IEEE transactions on affective computing},
  volume={13},
  number={3},
  pages={1195--1215},
  year={2020},
  publisher={IEEE},
  doi={10.1109/TAFFC.2020.2981446}
}

@article{mende2020delaware,
  title={The Delaware pain database: A set of painful expressions and corresponding norming data},
  author={Mende-Siedlecki, Peter and Qu-Lee, Jennie and Lin, Jingrun and Drain, Alexis and Goharzad, Azaadeh},
  journal={Pain reports},
  volume={5},
  number={6},
  pages={e853},
  year={2020},
  publisher={LWW},
  doi={10.1097/PR9.0000000000000853}
}

@book{rabiner1978digital,
  title={Digital processing of speech signals},
  author={Rabiner, Lawrence R},
  year={1978},
  publisher={Pearson Education India}
}

@article{oliva2020core,
  title={Core muscle activity during physical fitness exercises: A systematic review},
  author={Oliva-Lozano, Jos{\'e} M and Muyor, Jos{\'e} M},
  journal={International journal of environmental research and public health},
  volume={17},
  number={12},
  pages={4306},
  year={2020},
  publisher={MDPI},
  doi={10.3390/ijerph17124306}
}

@article {LSL,
	author = {Kothe, Christian and Shirazi, Seyed Yahya and Stenner, Tristan and Medine, David and Boulay, Chadwick and Grivich, Matthew I. and Mullen, Tim and Delorme, Arnaud and Makeig, Scott},
	title = {The Lab Streaming Layer for Synchronized Multimodal Recording},
	elocation-id = {2024.02.13.580071},
	year = {2024},
	doi = {10.1101/2024.02.13.580071},
	publisher = {Cold Spring Harbor Laboratory},
	URL = {https://www.biorxiv.org/content/early/2024/02/14/2024.02.13.580071},
	eprint = {https://www.biorxiv.org/content/early/2024/02/14/2024.02.13.580071.full.pdf},
	journal = {bioRxiv}
}

@article{gpaq,
  title = {Global Physical Activity Questionnaire (GPAQ): Nine Country Reliability and Validity Study},
  volume = {6},
  ISSN = {1543-5474},
  url = {http://dx.doi.org/10.1123/jpah.6.6.790},
  DOI = {10.1123/jpah.6.6.790},
  number = {6},
  journal = {Journal of Physical Activity and Health},
  publisher = {Human Kinetics},
  author = {Bull,  Fiona C. and Maslin,  Tahlia S. and Armstrong,  Timothy},
  year = {2009},
  month = nov,
  pages = {790–804}
}

@article{MET,
  title = {Metabolic equivalents (METS) in exercise testing,  exercise prescription,  and evaluation of functional capacity},
  volume = {13},
  ISSN = {1932-8737},
  url = {http://dx.doi.org/10.1002/clc.4960130809},
  DOI = {10.1002/clc.4960130809},
  number = {8},
  journal = {Clinical Cardiology},
  publisher = {Wiley},
  author = {Jetté,  M. and Sidney,  K. and Bl\"{u}mchen,  G.},
  year = {1990},
  month = aug,
  pages = {555–565}
}

@article{babb1991lung,
  title={Lung volumes during low-intensity steady-state cycling},
  author={Babb, TG and Rodarte, JR},
  journal={Journal of applied physiology},
  volume={70},
  number={2},
  pages={934--937},
  year={1991},
  doi={10.1152/jappl.1991.70.2.934}
}

@article{matthews1989exercise,
  title={Exercise responses during incremental and high intensity and low intensity steady state exercise in patients with obstructive lung disease and normal control subjects},
  author={Matthews, Joseph L and Bush, Bruce A and Ewald, Frank W},
  journal={Chest},
  volume={96},
  number={1},
  pages={11--17},
  year={1989},
  publisher={Elsevier},
  doi={10.1378/chest.96.1.11}
}

@inproceedings{lugaresi2019mediapipe,
  title={Mediapipe: A framework for perceiving and processing reality},
  author={Lugaresi, Camillo and Tang, Jiuqiang and Nash, Hadon and McClanahan, Chris and Uboweja, Esha and Hays, Michael and Zhang, Fan and Chang, Chuo-Ling and Yong, Ming and Lee, Juhyun and others},
  booktitle={Third workshop on computer vision for AR/VR at IEEE computer vision and pattern recognition (CVPR)},
  volume={2019},
  year={2019},
  doi={10.48550/arXiv.1906.08172}
}

@article{jonhagen2009forward,
  title={Forward lunge: a training study of eccentric exercises of the lower limbs},
  author={J{\"o}nhagen, Sven and Ackermann, Paul and Saartok, T{\"o}nu},
  journal={The Journal of Strength \& Conditioning Research},
  volume={23},
  number={3},
  pages={972--978},
  year={2009},
  publisher={LWW},
  doi={10.1519/JSC.0b013e3181a00d98}
}

@article{kidgell2010neurophysiological,
  title={Neurophysiological responses after short-term strength training of the biceps brachii muscle},
  author={Kidgell, Dawson J and Stokes, Mark A and Castricum, Troy J and Pearce, Alan J},
  journal={The Journal of Strength \& Conditioning Research},
  volume={24},
  number={11},
  pages={3123--3132},
  year={2010},
  publisher={LWW},
  doi={10.1519/JSC.0b013e3181f56794}
}

@article{tong2014sport,
  title={Sport-specific endurance plank test for evaluation of global core muscle function},
  author={Tong, Tom K and Wu, Shing and Nie, Jinlei},
  journal={Physical Therapy in Sport},
  volume={15},
  number={1},
  pages={58--63},
  year={2014},
  publisher={Elsevier},
  doi={10.1016/j.ptsp.2013.03.003}
}

@article{iglesias2010analysis,
  title={Analysis of factors that influence the maximum number of repetitions in two upper-body resistance exercises: curl biceps and bench press},
  author={Iglesias, Eliseo and Boullosa, Daniel A and Dopico, Xurxo and Carballeira, Eduardo},
  journal={The Journal of Strength \& Conditioning Research},
  volume={24},
  number={6},
  pages={1566--1572},
  year={2010},
  publisher={LWW},
  doi={10.1519/JSC.0b013e3181d8eabe}
}

@article{marchetti2018balance,
  title={Balance and lower limb muscle activation between in-line and traditional lunge exercises},
  author={Marchetti, Paulo H and Guiselini, Mauro A and da Silva, Josinaldo J and Tucker, Raymond and Behm, David G and Brown, Lee E},
  journal={Journal of human kinetics},
  volume={62},
  pages={15},
  year={2018},
  doi={10.1515/hukin-2017-0174}
}

@inproceedings{kirillov2023segment,
  title={Segment anything},
  author={Kirillov, Alexander and Mintun, Eric and Ravi, Nikhila and Mao, Hanzi and Rolland, Chloe and Gustafson, Laura and Xiao, Tete and Whitehead, Spencer and Berg, Alexander C and Lo, Wan-Yen and others},
  booktitle={Proceedings of the IEEE/CVF international conference on computer vision},
  pages={4015--4026},
  year={2023},
  doi={10.48550/arXiv.2304.02643}
}

@article{ai2023artificial,
  title={Artificial intelligence risk management framework (AI RMF 1.0)},
  author={AI, NIST},
  journal={URL: https://nvlpubs. nist. gov/nistpubs/ai/nist. ai},
  pages={100--1},
  year={2023},
  doi={https://doi.org/10.6028/NIST.AI.100-1}
}

@article{seo2024evaluation,
  title={Evaluation framework of large language models in medical documentation: Development and usability study},
  author={Seo, Junhyuk and Choi, Dasol and Kim, Taerim and Cha, Won Chul and Kim, Minha and Yoo, Haanju and Oh, Namkee and Yi, YongJin and Lee, Kye Hwa and Choi, Edward},
  journal={Journal of Medical Internet Research},
  volume={26},
  pages={e58329},
  year={2024},
  publisher={JMIR Publications Toronto, Canada},
  doi={10.2196/58329}
}

@baothesis{bao2022thesis,
  author       = {Bao, Ngo Hong Quoc},
  title        = {Wrong Pose Detection in Exercise Videos Based on Machine Learning},
  school       = {Can Tho University},
  year         = {2022},
  note         = {Accessed from Scribd, GitHub},
  url          = {https://github.com/NgoQuocBao1010/Exercise-Correction/tree/main},
}

@article{wang2024ubiphysio,
  title={Ubiphysio: Support daily functioning, fitness, and rehabilitation with action understanding and feedback in natural language},
  author={Wang, Chongyang and Feng, Yuan and Zhong, Lingxiao and Zhu, Siyi and Zhang, Chi and Zheng, Siqi and Liang, Chen and Wang, Yuntao and He, Chengqi and Yu, Chun and others},
  journal={Proceedings of the ACM on Interactive, Mobile, Wearable and Ubiquitous Technologies},
  volume={8},
  number={1},
  pages={1--27},
  year={2024},
  publisher={ACM New York, NY, USA}
}

@inproceedings{jorke2025gptcoach,
  title={GPTCoach: towards LLM-based physical activity coaching},
  author={J{\"o}rke, Matthew and Sapkota, Shardul and Warkenthien, Lyndsea and Vainio, Niklas and Schmiedmayer, Paul and Brunskill, Emma and Landay, James A},
  booktitle={Proceedings of the 2025 CHI conference on human factors in computing systems},
  pages={1--46},
  year={2025}
}

\end{document}